\documentclass{aa}
\usepackage{epsf}
\usepackage{graphicx}
\usepackage{natbib}
\usepackage{txfonts}
\usepackage{longtable}

\newcommand{\revised}{}

\def\d{$^{\rm o}$}
\def\vsini{$V\!\sin i$}
\def\teff{T$_{\rm eff}$}

\def\logg{$\log~g$}
\def\om{$\Omega$}
\def\omc{$\Omega/\Omega_{\rm{c}}$}
\def\omcd{$\Omega_{\rm{c}}$}
\def\kps{km~s$^{-1}$}

\def\top{$T_{\rm eff}^{\rm o}$}
\def\gop{$\log g_{\rm o}$}
\def\vsinio{$V\!\sin i_{\rm~\!true}$}

\begin{document}
\title{Fundamental parameters of Be stars\\
located in the seismology fields of COROT
\thanks{Based on GAUDI, the
data archive and access system of the ground-based
asteroseismology programme of the COROT mission. The GAUDI system
is maintained at LAEFF (http://ines.laeff.esa.es/corot/). LAEFF
is part of the Space Science Division of INTA. Also based on OHP
(Observatoire de Haute-Provence, France), LNA (Laborat\'orio
Nacional de Astrof\'{\i}sica, Brazil) and FEROS (ESO, {\revised
072.D-0315(A)}) observations. {\revised A part of the Tables and
Figures described in this paper are available at the CDS (Centre
de Donn\'ees astronomiques de Strasbourg:
http://cdsweb.u-strasbg.fr/cats/cats.html).}}}

\author{Y. Fr\'emat\inst{1}
        \and C. Neiner\inst{2,3}
        \and A.-M. Hubert\inst{3}
        \and M. Floquet\inst{3}
        \and J. Zorec\inst{4} \and\\
         E. Janot-Pacheco\inst{5}
        \and J. Renan de Medeiros\inst{6}
}

\offprints{Y. Fr\'emat,
\email{yves.fremat@oma.be}}


\institute{Royal Observatory of Belgium, 3 avenue circulaire, 1180
Brussels, Belgium \and Institute for astronomy, KU Leuven,
Celestijnenlaan 200B, 3001 Leuven, Belgium \and GEPI / UMR 8111 du
CNRS, Observatoire de Paris-Meudon, 5 place Jules Janssen, 92195
Meudon, France \and Institut d'Astrophysique de Paris, UMR 7095
CNRS, Universit\'e Pierre \& Marie Curie \and Instituto de
Astronomia, Geof\'{\i}sica e Ci\^encias Atmosf\'ericas,
Universidade de S\~ao Paulo, Rua do Mat\~ao 1226, 05508-090 S\~ao
Paulo, Brazil \and Departamento de Fisica, Universidade Federal
do Rio Grande do Norte, 59072-970 Natal, Brazil}

\titlerunning{Be stars in the field of COROT}
\authorrunning{Fr\'emat et al.}

\date{Accepted date: October 27, 2005}

\abstract{In preparation for the COROT space mission, we
determined the fundamental parameters (spectral type,
temperature, gravity, \vsini) of the Be stars observable by COROT
in its seismology fields (64 Be stars). We applied a careful and
detailed modeling of the stellar spectra, taking into account the
veiling caused by the envelope, as well as the gravitational
darkening and stellar flattening due to rapid rotation.
Evolutionary tracks for fast rotators were used to derive stellar
masses and ages. The derived parameters will be used to select Be
stars as secondary targets (i.e. observed for 5 consecutive
months) and short-run targets of the COROT mission. {\revised
Furthermore, we note that the main part of our stellar sample is
falling in the second half of the main sequence life time, and
that in most cases the luminosity class of Be stars is inaccurate
in characterizing their evolutionary status.}.  }

\maketitle

\section{Introduction}

\subsection{Be stars}

{\revised Be stars are non-supergiant B stars that show or have
shown at one or another moment H$\alpha$ emission
\citep{bestars}. More generally, emission does not only occur in
the first members of the Balmer line series, but can also affect
the continuum and line profiles of other atoms or ions, such as
\ion{Fe}{ii}. It is generally agreed that, in classical Be stars,
this emission is due to the presence of a cool, disk-like
circumstellar envelope concentrated in the equatorial plane.} For
a complete review of the "Be phenomenon" and its characteristics,
see \citet{2003PASP..115.1153P}.

{\revised Classical Be stars are fast rotators.} Statistical
studies indeed suggest that their angular speed ranges from 60 to
100\% of the critical breakup velocity ($\Omega_{\rm{c}}$), with
a narrow maximum occurrence at 80\% \citep{2001A&A...378..861C}
or at 90\% when accounting for fast rotation effects
\citep{effets} {\revised in order to limit the saturation of the
lines' FWHM at high rotation rates \citep{owocki}.}

Although we know that the envelopes of {\revised classical} Be
stars are probably formed during episodes of strong mass
ejection, the precise origin of these ejections is still unknown
{\revised and may differ from star to star}. One of the currently
most interesting explanations resides in the fast (non-critical)
rotation and the beating of non-radial pulsation (NRP) modes
combining their effects to cause matter expulsion. In the H-R
diagram, early Be stars are indeed located at the lower border of
the instability strip of the $\beta$\,Cep stars, while mid and
late Be stars are mixed with Slowly Pulsating B (SPB) stars.
Short and long period pulsations are therefore expected and
confirmed in the majority of Be stars, if we except the late
sub-classes. From the Hipparcos database
\citep{1998A&A...335..565H} it was shown that short-term
variability is present in almost all early-Be stars (86\%), while
it seems to be less common in the cooler ones (40\% to 18\% from
the mid to the latest spectral types). This fact, however, could
be due to the variability detection level of current
instrumentation, since the amplitude of pulsations in late
subtypes of B stars is expected to be very small (a few mmag). Up
to now, about hundred Be stars have already been claimed as
short-term periodic variables (hours, tens of hours) and their
number is still increasing. However, the detection of short and
long-term pulsations is difficult using ground-based observations
even in the framework of multisite campaigns \citep[e.g. the
Musicos 98 campaign, see][]{2002A&A...388..899N}. The prediction
and the study of the coincidence between the beating of
multiperiodic pulsations and the occurrence of mass-ejection
further needs the determination of accurate periods and the
detection of the most complete sample of pulsation modes,
{\revised which is not an easy matter, even during multisite
campaigns. Although multiperiodicity has been detected in several
Be stars \citep[e.g. 66 Oph, ][]{floquet2002}, only one such
coincidence case between beating and matter ejection has been
reported until now} \citep[$\mu$\,Cen,][]{1998A&A...336..177R}.
{\revised As a matter of fact, most of the presently known cases
of pulsation in Be stars result from high-resolution spectroscopy.
This technique is very powerful in identifying high-order
pulsation modes, but not well adapted to resolve closely spaced
multiple periods. Photometry with COROT will therefore provide a
much more superior observing tool to detect new multiperiodic
pulsators among Be stars.}

{\revised In this framework,} the COROT mission will provide us
with 5-months continuous observations of a substantial amount of
Be stars. It will allow us to reach a never achieved precision in
the determination of pulsation frequencies, further giving us the
opportunity to study their possible magnetic splitting and the
rotational modulation in Be stars.

\subsection{The COROT mission}

{\revised The COROT (COnvection, ROtation and planetary Transits)
satellite\footnote{http://corot.oamp.fr/} will be launched in
August 2006 and has two goals: to study the interior of stars by
looking at their oscillations and to search for extrasolar
planets by detecting planetary transits
\citep[see][]{2002ASPC..259..626B}. Therefore four CCDs are used:
two for the asteroseismology program and two for the exoplanetary
program. The asteroseismology CCDs are positioned on the sky to
observe simultaneously one (or sometimes two) bright (V $\sim$
5.5) primary target(s) plus eight or nine surrounding secondary
targets with a magnitude 5.6 $\le$ V $\le$ 9.4.

The 30 cm telescope of COROT will be pointed alternatively towards the galactic
centre and anticentre. These two cones of observations, which have a radius of
10 degrees, are the pointing limits for the CCDs. The two asteroseismology CCDs
cover a field of view of 1.3 $\times$ 2.6 degrees. At least 5 such fields will
be observed by COROT during $\sim$5 months each.  Finally, short observing runs
(20-30 days) will be performed on specific targets so that the full target list
covers as well as possible the HR diagram.

Bright Be stars can therefore be observed either as secondary
targets of the asteroseismology program in long runs, or as short
runs targets.  An international collaboration led by A.-M. Hubert
is preparing these observations. In this paper, we present the
procedure that was carried out to perform a preliminary analysis
of as many bright (5.6 $<$ V $<$ 9.4) classical Be stars as
possible in the observing cones of COROT. The position, spectral
type and variability of these Be stars are factors taken into
account for the selection of the fields that will be observed by
COROT. To carry out this first target selection, we therefore
need good stellar parameters with typical accuracies of 10 \% on
the effective temperature (\teff), of 0.1 to 0.2 dex on the
surface gravity (\logg), and of 5 to 10 \% on the projected
rotation velocity (\vsini). To be valuable, the parameters
determination further needs to account for the peculiar nature of
Be stars, including the effects of fast rotation, circumstellar
emission, and departure from LTE.}

Most of the data used in this study are available on the GAUDI
{\revised (Ground-based Asteroseismology Uniform Database
Interface)} database \citep{solano} and are described in
Sect~\ref{sect:observations}. In GAUDI, more specific
informations, such as fundamental stellar parameters (surface
gravity and effective temperature) and projected rotation
velocities, can also be found and were automatically derived from
the observations. However, these parameters were obtained without
accounting for neither circumstellar emission nor fast rotation
effects, which are generally expected to affect the spectra of Be
stars. The goal of the present work is therefore to reestimate
the effective temperature (\teff), surface gravity (\logg) and
projected rotation velocity (\vsini) accounting, as far as
possible, for the peculiar nature of Be stars, {\revised in order
to update the GAUDI database and thus facilitate the target
selection for COROT.

For the same reasons, this work also aims at identifying the
stars in our sample that could be considered as {\it particular,}
such as Herbig stars or spectroscopic binaries with a Be star
component.} The model atmospheres we used are defined in
Sect.~\ref{sec:modeles}, while the procedure we adopted to
perform these determinations is detailed in
Sect.~\ref{sect:adopted_procedure}. Our results are listed in
Sect.~\ref{sec:results}, with remarks concerning several specific
targets gathered in Sect.~\ref{sec:remarks}, and discussed in
Sect.~\ref{sec:discussion}.

\begin{table}
\caption[]{Sample of spectra used for this study. A complete
version of the table is available at the CDS {\revised
(http://cdsweb.u-strasbg.fr/cats/cats.html)}. {\revised For the
ELODIE spectra, the signal to noise ratio (S/N) was provided by
the INTERTACOS (OHP) reduction pipeline, while for the other data
it was computed with IRAF by selecting some parts of the
continuum in the studied spectral region.} '$^1$' indicates that
the spectrum is available in GAUDI. } \label{tab:obs}
\begin{center}
\begin{tabular}{@{}llll@{}r@{}}
\hline \hline \noalign{\smallskip}
HD & Obs. date & T$_{\rm exp}$ (s) & Instrument & S/N \\
\hline\noalign{\smallskip}
42406     & 2004-02-05 & 900      & FEROS &249\\  
43264     & 2001-11-27 & 3300  & ELODIE$^1$ &83\\
43285     & 2001-12-21 & 1800  & ELODIE$^1$ &123\\
44783     & 2000-12-18 & 1500  & ELODIE$^1$ &121\\
45901     & 2004-01-03 & 2700  & AURELIE 4280\_G3&120\\
46380     & 2001-12-22 & 3600  & ELODIE$^1$ &53\\
46484     & 2003-01-26 & 3600  & ELODIE$^1$ &107\\
47054     & 2002-01-28 & 300   & FEROS$^1$ &145\\
\multicolumn{4}{l}{... {\it A complete version of the table is
available at
the CDS}}\\
\hline\noalign{\smallskip}
\end{tabular}
\end{center}
\end{table}

\section{Observations}

\label{sect:observations}

In preparation for the COROT satellite observations, an ambitious
ground-based observing program was performed (P.I.: C. Catala,
Observatoire de Paris). For each star with a magnitude between
{\revised 5.6} and 8 located in the observing cones of the COROT
satellite, at least one spectrum was obtained {and stored in the
GAUDI database}. {\revised The sample of stars we are studying in
the present paper is therefore a sub-sample of the compilation by
\citet{solano}. It gathers the targets that are well known Be
stars or that were recently identified as Be stars
\citep{neiner2005}}.

The data were mainly obtained with two high-resolution \'echelle
spectrographs (R $\sim$ 40000 -- 50000): ELODIE at the 2m
telescope of the Observatoire de Haute-Provence (OHP, France) and
FEROS at the 1.5m and 2.2m telescopes of ESO (La Silla, Chile).
Additional observations were also obtained at the 1.9 meter
telescope at SAAO (South Africa) with the GIRAFFE spectrograph,
with the CORALIE spectrograph on 1.2 m Swiss telescope in La
Silla (Chile), with the SARG ({\revised Spectrografo Alta
Risoluzione Galileo}) spectrograph at the 3.6m Telescopio
Nazionale Galileo (TNG, La Palma, Spain) and with the Coud\'e
spectrograph on the 2 meter telescope of the Tautenburg
observatory (Germany).


{\revised To complete the sample of Be stars in the observing
cones of COROT, additional spectra of Be stars with a magnitude
between 8 and 9.4 were obtained:} in Brazil, at the LNA
(Laborat\'orio Nacional de Astrofisica, Observatório do Pico dos
Dias) with the Cassegrain spectrograph {\revised (OPD CASS)}
attached to the 1.6 m Boller \& Chivens telescope and using the
900 lines/mm grating (R $\sim$ 7000); in France, at the OHP
(Observatoire de Haute-Provence) with AURELIE (1.52-m telescope)
at medium resolution (R $\sim$ 15000 with the grating N$^o$2,
$\Delta\lambda$~=~220~\AA) and lower resolution (R $\sim$ 7000
with the grating N$^o$3, $\Delta\lambda$ = 440 \AA); and in Chile
with FEROS (ESO) on Brazilian time.

{\revised In principle, the sample of studied stars contains all
Be stars in the observing cones of COROT with the adequate
magnitude (5.6 $\le$ V $\le$ 9.4). The target list was compiled
using SIMBAD and adding newly discovered Be stars from
\cite{neiner2005}. However, known SB2 with a Be star component
and interacting Be binaries were rejected from the list of
possible Be targets. A few faint Be stars are also not studied in
this paper because no spectra were obtained due to bad weather
conditions. As a consequence, the sample of studied Be stars is
not complete, but no systematic bias is expected. In particular,
the samples in the centre and anticenter direction are
equivalent.}

\section{Model atmospheres and flux grids}

\label{sec:modeles}

The plane-parallel atmosphere models we used for effective
temperatures ranging from 15000 K to 27000 K were computed in two
consecutive steps. To account in the most effective way for
line-blanketing, the temperature structure of the atmospheres was
computed by \citet{cdrom13} using the ATLAS9 FORTRAN program.
Non-LTE level populations were then calculated for each of the
atoms we considered using {\sc tlusty}
\citep{1995ApJ...439..875H} and keeping fixed the temperature and
density distributions. The surface chemical abundances we adopted
are those published by \citet{1998SSRv...85..161G} for the Sun.

Table \ref{tab:desc} lists the ions we introduced in the
computations. Except for C~{\sc ii}, the atomic models we used in
this work were downloaded from {\sc tlusty}'s
homepage\footnote{http://tlusty.gsfc.nasa.gov} maintained by I.
Hubeny and T. Lanz. C {\sc ii} was treated thanks to the {\sc
modion} IDL package developed by \citet{modion} and by adopting
the atomic data (oscillator strengths, energy levels and
photoionization cross sections) selected from the {\sc topbase}
database \citep{1993A&A...275L...5C}. It reproduces the results
obtained by \citet{1996ApJ...473..452S}.

\begin{table}[ht]
\caption{Atoms and ions treated in our computations assuming
NLTE. The number of levels taken into account for each ion is
also given.} \label{tab:desc} \center
\begin{tabular}{lll}
\hline
\hline
Atom & Ion & Levels\\
\hline \noalign{\smallskip}
Hydrogen & H {\sc i} & 8 levels + 1 superlevel\\
         & H {\sc ii} & 1 level\\
Helium   & He {\sc i} & 24 levels\\
         & He {\sc ii} & 20 levels\\
         & He {\sc iii} & 1 level\\
Carbon   & C {\sc ii} & 53 levels all individual levels\\
         & C {\sc iii} & 12 levels\\
         & C {\sc iv}  & 9 levels + 4 superlevels\\
         & C {\sc v} & 1 level\\
Nitrogen & N {\sc i} & 13 levels\\
         & N {\sc ii} & 35 levels + 14 superlevels\\
         & N {\sc iii} & 11 levels\\
         & N {\sc iv} & 1 level\\
Oxygen   & O {\sc i} & 14 levels + 8 superlevels\\
         & O {\sc ii} & 36 levels + 12 superlevels\\
         & O {\sc iii} & 9 levels\\
         & O {\sc iv} & 1 level\\
Magnesium & Mg {\sc ii} & 21 levels + 4 superlevels\\
          & Mg {\sc iii} & 1 level\\
\hline
\end{tabular}
\end{table}

Model atmospheres with effective temperatures lower than 15000 K
were treated assuming full LTE, while those hotter than 27000 K
were taken from the OSTAR2002 NLTE grid
\citep{2003ApJS..146..417L}. The grid of fluxes we use during the
fitting procedure (Sect.~\ref{sed:proc:fund}) was finally built
with {\sc synspec} and for effective temperatures and surface
gravities ranging from 8000 to 50000 K and from 2.5 to 4.5 dex
(cgs) respectively.

\section{Adopted procedure}

\label{sect:adopted_procedure}

To take into account the main phenomena expected to affect the
spectra of Be stars, the determination of their fundamental
parameters was carried out in three consecutive steps, each of
them being described in the following sections and summarized in
Fig.~\ref{fig:veil}.

\begin{figure}
\center
\includegraphics[width=8cm,clip=]{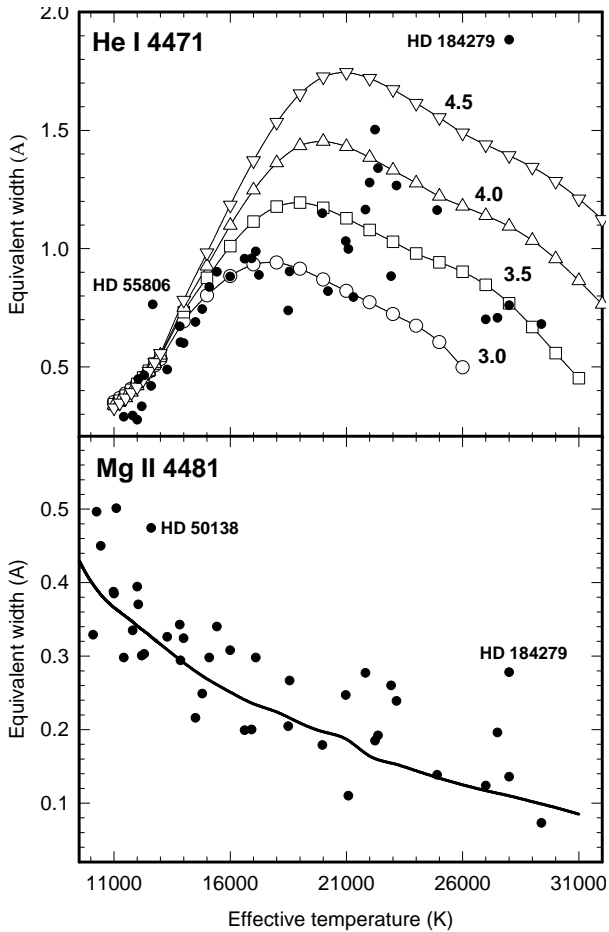}
\caption{Computed (curves) and observed (filled circles)
equivalent widths for the \ion{He}{i} 4471 (upper panel) and
\ion{Mg}{ii} 4481 (lower panel) spectral lines are shown. The
surface gravity adopted in the calculations is noted on each
curve. \ion{Mg}{ii} 4481 has a very weak luminosity-dependence,
for clarity we therefore plotted only the theoretical line
corresponding to \logg~=~4.0. {\revised Computations are made
using plane-parallel model atmospheres (Sect. 3).}}
\label{fig:eqw}
\end{figure}

\subsection{Apparent fundamental parameters determination}

\label{sed:proc:fund}

\begin{figure*}
\center
\includegraphics[width=17cm,clip=]{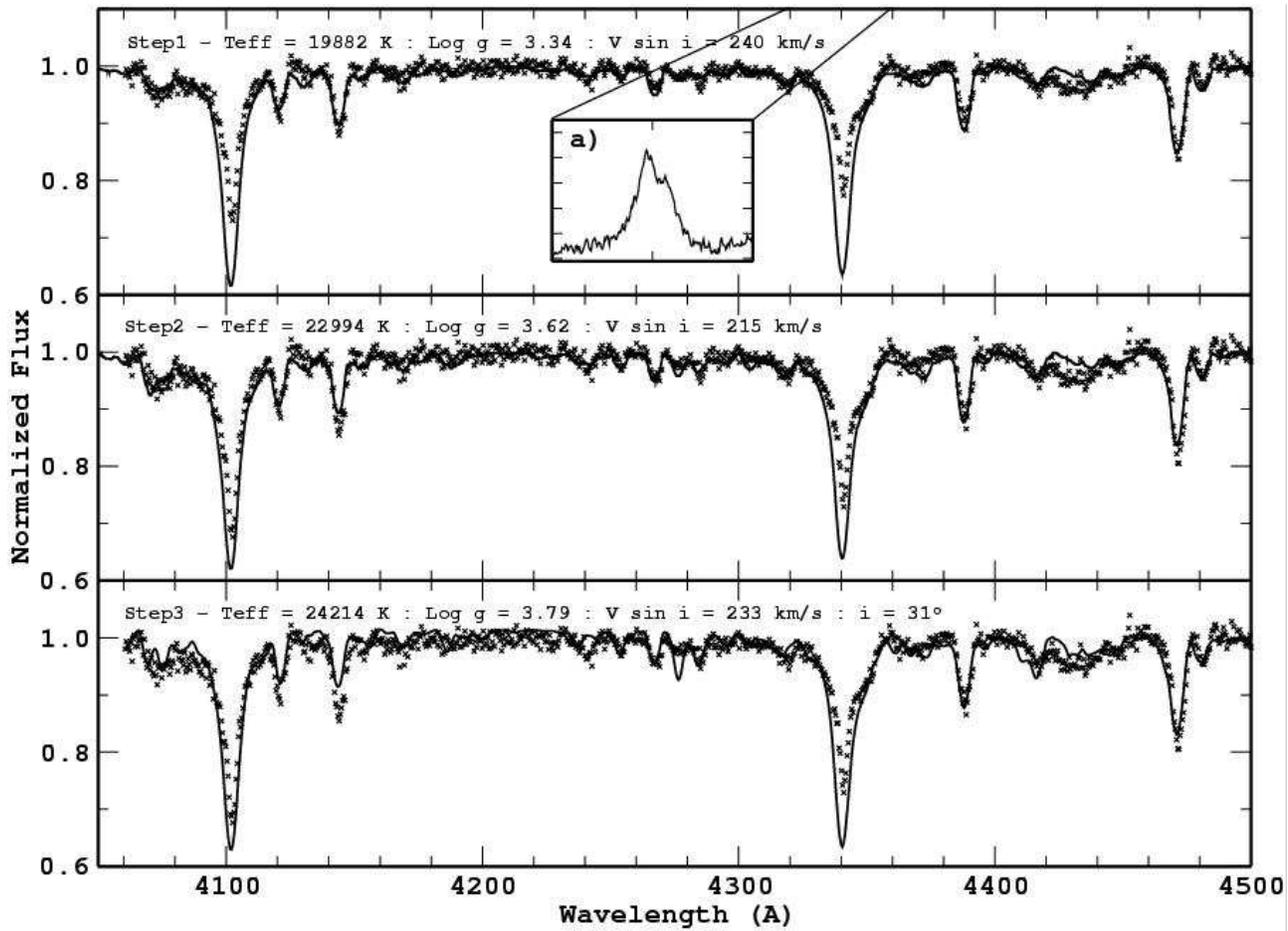}
\caption{Description of the three-steps procedure adopted to
derive the fundamental parameters. {\bf Step 1:} Fit of the
observed spectrum (crosses) with synthetic spectra (solid line),
ignoring continuum veiling and second order fast-rotation
effects. Subtracting the observed H$\gamma$ line-profile from the
synthetic one allows us to estimate the magnitude of the
circumstellar emission (panel a). {\bf Step 2:} Observations are
corrected for veiling {\revised and fitted again}. {\bf Step 3:}
The fundamental parameters derived in step 2 are corrected to
account for fast rotation effects {\revised with a new fit} at
\omc=0.99.} \label{fig:veil}
\end{figure*}

Hydrogen, helium and \ion{Mg}{ii} lines are generally assumed to
be good temperature and gravity indicators for the study of
B-type stars. Fig.~\ref{fig:eqw} shows the variation of the
equivalent width computed for the \ion{He}{i} 4471 and
\ion{Mg}{ii} 4481 spectral lines with effective temperature and
surface gravity. Their different broadening mechanisms and
transition probabilities further present the advantage to have
been studied with great detail for a long time, allowing accurate
line profile computations. In our procedure, we therefore mainly
focus on a spectral domain ranging from 4000 to 4500~\AA~ (see
line-identification {\revised and computed equivalent widths} in
Table~\ref{tab:id}), which gathers no less than two hydrogen
lines, 5 strong helium lines and 2 blended \ion{Mg}{ii} lines.
Observations obtained in this region, and for each of the
considered Be stars, are compared to a grid of synthetic spectra
(see Steps 1 and 2 of Fig.~\ref{fig:veil}). For efficiency
reasons, this comparison is performed by means of a least squares
method based on the {\sc minuit} minimization package developed
at CERN, which we transposed to a FORTRAN computer code named
{\sc girfit}, and which allows to deal with large datasets. {\sc
girfit} interpolates the spectra in a grid of stellar fluxes
computed with plane-parallel model atmospheres (see
Sect.~\ref{sec:modeles}) for different values of the effective
temperature and of the surface gravity. To account for the
instrumental resolution and the Doppler broadening due to
rotation, the spectra are then convolved with a Gaussian function
using subroutines taken from the {\sc rotins} computer code
provided with {\sc synspec} \citep{1995ApJ...439..875H}. During
the fitting procedure, 5 free parameters are considered:
{\revised the effective temperature $T_{\rm eff}$, the surface
gravity $\log g$, the projected rotation velocity \vsini, the
radial velocity $V_{\rm rad}$ and the wavelength-independent ratio
between the mean "flux" level of the normalized observed and
theoretical spectra, i.e. a scaling factor allowing to match the
stellar continuum.} The $\chi^2$ parameter is computed on
different spectral zones chosen from 4000 to 4500~\AA. These
zones are selected in order to exclude any part of the spectrum
that could be affected by line emission or shell absorption (e.g.
hydrogen line cores) or/and by interstellar absorption bands
(generally found between 4400 and 4450 \AA). As the parameters
derived in this way do not take into account the effects of
stellar flattening and gravitational darkening due to fast
rotation, they will be further called {\it apparent} fundamental
parameters {\revised and correspond to what is obtained when
assuming that the star is a sphere with uniform temperature and
density surface distributions}.

\begin{table}
\caption{Theoretical equivalent widths of the main spectral lines
found in the considered spectral region and for various spectral
types. {\revised Computations are made using plane-parallel model
atmospheres (Sect. 3). Effective temperatures and spectral types
are taken from \citet{1994AJ....107..742G}.}} \label{tab:id}
\begin{center}
\begin{tabular}{llrrrr}
\hline \hline\noalign{\smallskip}
                   & & \multicolumn{4}{c}{Equivalent width (m\AA)} \\
\cline{3-6}
Ion                & $\lambda$ (\AA) & B0V &  B2V & B5V & B8V \\
 & & 29000 K & 19500 K & 14000 K & 11550 K\\
\hline\noalign{\smallskip}
\ion{He}{i}        &  4009 & 334  & 613  & 217  & 48    \\
\ion{He}{i}        &  4026 & 908  & 1541 & 878  & 542    \\
\ion{H}{i}         &  4102 & 3869 & 6452 & 9358 & 11475  \\
\ion{He}{i}        &  4121 & 450  & 251  & 82   & 32       \\
\ion{He}{i}        &  4144 & 443  & 765  & 322  & 120    \\
\ion{He}{i}        &  4169 & 88   & 91   & 32   &  5     \\
\ion{C}{ii}        &  4267 & 215  & 270  & 97   &  49    \\
\ion{H}{i}         &  4340 & 3724 & 6368 & 9552 & 12032 \\
\ion{He}{i}        &  4388 & 537  & 950  & 378  &  220      \\
\ion{He}{i}        &  4471 & 1087 & 1442 & 667  & 345    \\
\ion{Mg}{ii}       &  4481 & 122  & 198  & 272  & 318    \\
\hline
\end{tabular}
\end{center}
\end{table}


\subsection{Veiling caused by the envelope}

\label{sec:proc:veil}

The spectra of Be stars are not only affected by line-emission,
but they are also proportionally affected by continuum emission
and electron scattering, which change the stellar continuum
level. When significant, emission or/and scattering cause an
artificial weakening of the spectral line intensity generally
leading to underestimation of the effective temperature and of
the surface gravity in B-type stars. \citet{1995A&AS..111..423B}
estimated the magnitude of this veiling by studying hydrogen and
helium lines and proposed an empirical approach to correct the
stellar spectra from its effects. They mainly introduce a
correction term, r, which they found directly proportional to the
intensity or to the equivalent width of the H$\gamma$
line-emission W$_{\rm e}$(H$\gamma$) (see Step 1 of
Fig.~\ref{fig:veil}).

To measure the magnitude of the H$\gamma$ emission, the synthetic
H$\gamma$ profile obtained from a first fit of the fundamental
parameters was subtracted from the observations (see
Fig.~\ref{fig:veil}). The result was wavelength-integrated in
order to obtain W$_{\rm e}$(H$\gamma$), and to interpolate the r
value from Fig.~9a in \citet{1995A&AS..111..423B}. When greater
than zero, this correction was directly applied to the
observations \citep[see Eq. 1 in ][]{1995A&AS..111..423B}, which
were finally used to re-derive the fundamental parameters (see
Sect.~\ref{sed:proc:fund}).

\begin{figure*}
\center
\includegraphics[width=17cm,clip=]{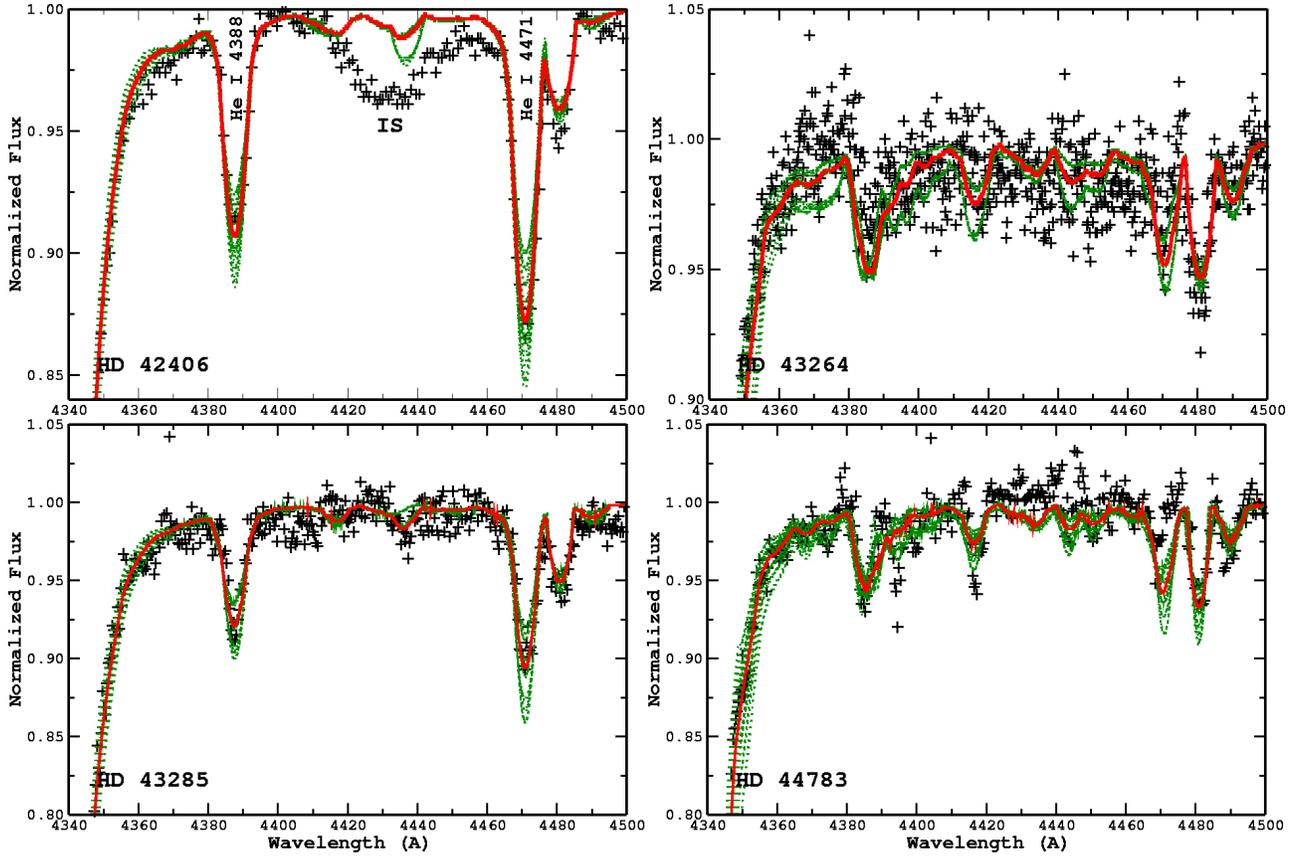}
\caption{Comparison between observed data (crosses) and fitted
synthetic spectra (full line) in a spectral domain containing the
red wing of the H$\gamma$ line and 2 neutral helium lines.
{\revised Dotted lines are used to represent 26 spectra computed
for different combinations (i.e. there are in fact $3^3$ spectra
or stellar parameters combinations, but one of them corresponds
to the best fit we obtained) of the upper and lower limits of the
\teff, \logg, and \vsini~values adopting the error bars given in
Table~\ref{tab:param}.}} \label{fig:aj}
\end{figure*}

\subsection{Gravitational darkening and stellar flattening}
\label{sec:gd}


As mentioned in the introduction, Be stars are fast rotators with
angular velocities probably around 90\% of their breakup
velocity. It is expected that a fast and solid-body-type rotation
flattens the star, causing a gravitational darkening of the
stellar surface due to the variation of the temperature and
density distribution from the poles to the equator. For Be stars,
we therefore have to account for these effects on the stellar
spectra and, consequently, on the determination of the
fundamental parameters. In the present paper, these effects are
introduced as corrections (see Step 3 of Fig.~\ref{fig:veil})
directly applied to the fundamental parameters derived in the two
previous steps (Sect.~\ref{sed:proc:fund} and
\ref{sec:proc:veil}). These corrections were estimated by
assuming different rotation rates and by systematically comparing
a grid of spectra taking into account the effects of fast
rotation \citep{2004IAUS..215...23F,effets} to a grid of spectra
computed using usual plane-parallel model atmospheres (see
Sect.~\ref{sec:modeles}). Eventually, the complete procedure
provides us with the parameters of the stellar {\it parent
non-rotating counterpart} (i.e. parameters that the stars would
have if they were rotationless), further called {\it pnrc}
parameters, estimated at different \omc\ values. It is the value
of these {\it pnrc} parameters that should be preferred to
interpret and discuss the future COROT data.

\begin{figure}
\center
\includegraphics[width=8cm,clip=]{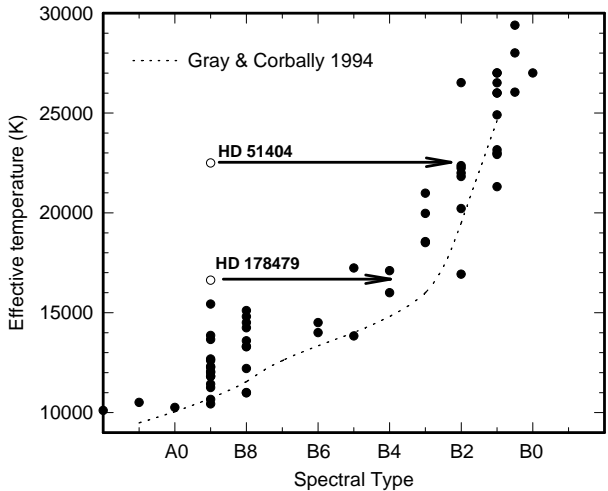}
\caption{{\revised Veiling corrected apparent} effective
temperatures listed in Table~\ref{tab:param} are reported versus
their spectral type given in SIMBAD (filled circles) and compared
to the calibration obtained by \citet{1994AJ....107..742G} for
dwarf stars (dashed line). \label{fig:cal}}
\end{figure}

\section{Results}

\label{sec:results}

The procedure described in Sect.~3 was applied to the sample of
selected Be stars (Table \ref{tab:obs}). {\revised We show in
Fig.~\ref{fig:aj} an example of observed and fitted-synthetic
spectra limited to a small part of the considered spectral domain.
A more complete comparison performed for all the studied stars is
available as a postscript file, which can be downloaded from the
CDS. In the same figure, we also plotted with dotted lines 26
synthetic spectra computed for different combinations (i.e. $3^3$
spectra or combinations, one of them corresponding to the best fit
we obtained) of the upper and lower limits of the \teff, \logg,
and \vsini~values adopting the error bars given in
Table~\ref{tab:param}.}

We plot in Fig.~\ref{fig:cal} the {\revised apparent value} of the
effective temperature we derived against the spectral type
available in the SIMBAD database. {\revised This enables us to
compare our determinations with older measurements and to detect
any inconsistency (i.e. HD 51404 and HD 184479). Such a
comparison is useful since, as far as the hydrogen and neutral
helium lines are considered, the apparent stellar parameters
characterize the spectrum fairly well \citep[e.g. see Fig. 9 in
][]{effets} and can therefore be directly related to the spectral
type given in SIMBAD.} Taking into account the fact that our
sample also includes stars that have luminosity classes generally
ranging from III to V, the distribution of points fairly follows
the effective temperature calibration proposed by
\citet{1994AJ....107..742G} for dwarf stars (dashed line in
Fig.~\ref{fig:cal}). Table~\ref{tab:param} gathers the derived
{\revised apparent} fundamental parameters: cols. 1 and 2
identify the target; cols. 3 and 4 respectively give the V
magnitude and spectral type extracted from the SIMBAD database;
cols. 5, 6, 7 and 8 list the effective temperature (\teff),
surface gravity (\logg), projected rotation velocity (\vsini) and
spectral type we obtained; values found in cols. 9 and 10 are
estimates of the H$\gamma$ line-emission and continuum veiling
due to the presence of the circumstellar envelope; col. 10
summarizes previous fundamental parameter determinations. The
spectral type we give in col. 8 is an estimate based on the
fundamental parameters combined to the \teff~and
\logg~calibrations proposed by \citet{1994AJ....107..742G} and by
\citet{1986serd.book.....Z}. Note that, at this stage, our
results do not account for stellar flattening and gravitational
darkening, but only for veiling. The derived stellar parameters
may therefore be considered as veiling-corrected {\it apparent}
values. {\revised We computed the luminosity of each target
combining these apparent \teff~and \logg~determinations to the
theoretical evolutionary tracks of \citet[][
Z=0.02]{1992A&AS...96..269S}. Their position in the HR diagram is
plotted in Fig.~\ref{fig:hres}a. Luminosity accuracy is estimated
from the \teff~and \logg~error-boxes.}

The magnitude of the uncertainty introduced by the fast-rotation
effects was estimated by applying the approach detailed in
Sect.~\ref{sec:gd} \citep[see also][]{effets} and by assuming
different \omc~ratios (where \om~and \omcd~respectively are the
actual and break-up angular velocities). Rotation corrected
fundamental parameters (\top, \gop~and \vsinio), i.e. $pnrc$
parameters, and corresponding inclination angles, $i$, are given
in Table~\ref{tab:gdcor} for \omc~=~0.8, 0.9, 0.95, and 0.99.

\begin{table*}[t]
\caption[]{\revised Veiling corrected apparent stellar parameters.
ID numbers, SIMBAD V magnitudes and spectral types are given for
each target in cols. 1, 2, 3 and 4. The derived stellar parameters
(effective temperature, surface gravity, and \vsini) are gathered
in cols. 5, 6 and 7. Their accuracy is estimated by scanning the
solutions space while adopting different initial values for the
parameters. Spectral types (col. 8) are derived from the apparent
stellar parameters combined to the \teff~and \logg~calibrations
proposed by \citet{1994AJ....107..742G} and by
\citet{1986serd.book.....Z}. Cols. 9 and 10 list the equivalent
width of the H$\gamma$ emission components and the estimates of
the veiling correction, respectively. The error bars on the
equivalent widths are generally of the order of 15\% and are a
product of the fitting process. Errors on the veiling parameter,
r, are estimated by accounting for the accuracy on
W$_{\lambda}^{\rm{H}\gamma}$ and by assuming a 95\% confidence
interval on the reference data of \citet[Fig.
9a,][]{1995A&AS..111..423B}. \label{tab:param} }
\begin{center}
\begin{tabular}{rrcllcllrlrrl}
\hline\hline
\noalign{\smallskip} \multicolumn{2}{c}{ID} && \multicolumn{2}{c}{SIMBAD} && \multicolumn{4}{c}{This work}\\
\cline{1-2}\cline{4-5}\cline{7-10} \noalign{\smallskip}
HD/BD &HIP/BD/&& V  & Sp. Type     && T$_{\rm eff}$ & \logg & V sin {\it i} & Sp. Type   & W$_{\lambda}^{\rm{H}\gamma}$     &   r  \\
      &  MWC  &&    &          && (K)       &    (c.g.s.) & (km s$^{\rm{-1}}$)  &     & (m\AA)     \\
\hline\noalign{\smallskip}
42406 & 29298      && 8.01  & B9        && 15400$\pm$1000 & 3.72$\pm$0.10  &  300$\pm$25 & B4 IV       & 0.30$\pm$0.05 & -- \\
43264 & 29719      && 7.51  & B9        && 10500$\pm${\revised 1000} & 2.76$\pm$0.10 & 288$\pm$10   & B9 III & 0.22$\pm$0.03 & --    \\
43285 & 29728      && 6.05  & B6Ve      && 14000$\pm$1000 & 3.78$\pm$0.10 & 260$\pm$20  & B5 IV& 0.21$\pm$0.03 & --     \\ 
44783 & 30448      && 6.225 & B8Vn      && 11000$\pm$1000 & 3.05$\pm$0.15 & 226$\pm$50 & B9 III & 0.10$\pm$0.02 & --      \\ 
45901 & 30992      && 8.87  & B2Ve      && 26500$\pm$2000 & 3.73$\pm$0.15 & 164$\pm$15 & B0.5 IV & 0.82$\pm$0.12 & 0.17$\pm$0.08 \\
\multicolumn{10}{l}{... {\it A complete version of the table is
available at the CDS}}\\
\hline
\end{tabular}
\end{center}
\end{table*}

\begin{table*}
\caption{Sample of fundamental parameters corrected for the
effects of fast rotation at different \omc~ratios. {\revised
Error bars on the parameters are of the same order as in
Table~\ref{tab:param}. When the projected rotation velocity is
greater than the break-up speed or equatorial speed, the Table is
left blank.} } \label{tab:gdcor} \center
\begin{tabular}{@{\ }ll@{\ \ }l@{\ \ }l@{\ }l@{\ \ }cl@{\ \ }l@{\
\ }l@{\ }l@{\ \ }cl@{\ \ }l@{\ \ }l@{\ }l@{\ \ }cl@{\ \ }l@{\ \
}l@{\ }l@{\ }} \hline \hline
   & \multicolumn{4}{c}{\omc~= 0.80} && \multicolumn{4}{c}{\omc~= 0.90} && \multicolumn{4}{c}{\omc~= 0.95} && \multicolumn{4}{c}{\omc~= 0.99}\\
\cline{2-5}\cline{7-10}\cline{12-15}\cline{17-20}
HD &  \top & \gop & \vsinio & $i$ && \top & \gop & \vsinio & $i$ && \top & \gop & \vsinio & $i$ && \top & \gop & \vsinio & $i$\\
   &   (K)  & (cgs) & (\kps) & ($^{\rm o}$) && (K) & (cgs) & (\kps) & ($^{\rm o}$) && (K)  & (cgs) & (\kps) & ($^{\rm o}$) && (K) & (cgs) & (\kps) & ($^{\rm o}$)\\
\hline\noalign{\smallskip}
 42406 &        &       &       &     &&        &       &       &     && 16500  &  3.96 &  338  &  61 && 17000  &  3.97 &  360  & 79\\
 43264 &        &       &       &     &&        &       &       &     &&        &       &       &     && 12000  &  3.16 &  284  &  79 \\
 43285 & 15000  &  4.03 &  266  &  71 && 15000  &  3.99 &  274  &  55 && 15000  &  3.95 &  292  &  51 && 15000  &  3.93 &  309  &  58 \\
 44783 & 12000  &  3.37 &  226  &  93 && 12000  &  3.40 &  227  &  80 && 11500  &  3.13 &  231  &  82 && 11500  &  3.21 &  218  &  57\\
 45901 & 27000  &  3.78 &  171  &  31 && 27000  &  3.79 &  173  &  26 && 27500  &  3.84 &  175  &  24 && 29500  &  4.06 &  174  &  21 \\
\multicolumn{10}{l}{... {\it A complete version of the table is available at the CDS}}\\
\hline
\end{tabular}
\end{table*}



\section{Remarks about specific targets}
\label{sec:remarks}



\subsection{HD 43264}

The Be nature of HD 43264 was first noted by \cite{neiner2005}.
However, the surface gravity we derive from the study of the
hydrogen and helium lines is quite low (\logg~=~2.76), which
means that the target could be a bright giant. The star is known
as a binary in the HIPPARCOS catalogue, probably SB1 considering
the magnitude difference between the primary and secondary
components ($\Delta$V~=~V{\sc a}~--~V{\sc b}~=~-2.85).

\subsection{HD 46380, HD 50087}

The observed spectra we used to derive the fundamental parameters of HD~46380
and HD~50087 are very noisy, and thus the parameters are uncertain.

\subsection{HD 50138}

HD 50138 has a variable shell and is also considered as a B[e]
star, whose evolutionary status is very difficult to establish.
Several hints \citep{1998A&A...340..117L} suggest that the star
could probably be a massive Herbig Be star with an accreting
circumstellar disk, i.e. in a pre-main sequence. The spectrum is
very complex with strong emission lines and thin absorption
features superimposed on the photospheric lines. The
determination of fundamental parameters from the fit of the
spectrum is therefore very difficult and could be inaccurate.

\subsection{HD 51404}

HD~51404 is a poorly studied object recognized as Be star by
\citet{1949ApJ...110..387M} and erroneously classified as a B9 V
star. Our determinations clearly show that its spectral type is
B1.5, as can be deduced from Fig.~\ref{fig:cal} and from the
strength of the \ion{He}{i} spectral lines.

\subsection{HD 52721}

HD~52721 is also known as a Herbig Ae/Be candidate \citep{2003AJ....126.2971V}
and is a member of a visual double system with angular separation $\sim$ 0.65
arcsec \citep{1997A&A...323L..49P}. Though a fair agreement between observed
and theoretical fitted spectra is observed, the \vsini~value we derive (352
\kps) strongly deviates from those obtained in previous works: 243 $\pm$ 93
\kps \citep{2001A&A...368..912Y} and 456 \kps \citep{1996PASP..108..833H}.

\begin{figure}
\center
\includegraphics[width=8cm,clip=]{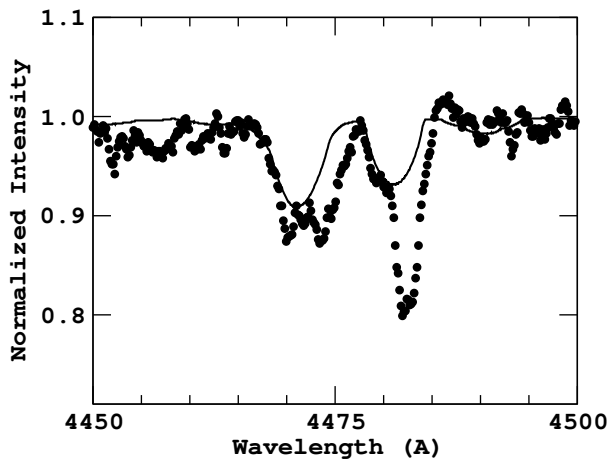}
\caption{Observed (dots) and fitted (solid line) \ion{He}{i} 4471
and \ion{Mg}{ii} 4481 lines in the spectrum of HD 55806.}
\label{fig:55806}
\end{figure}

\subsection{HD 55806}

\label{sec:55806}

HD~55806 is a poorly studied B type star that was found to have bright
emission lines by \citet{1949ApJ...110..387M}. We were not able to find any
set of fundamental parameters allowing to simultaneously fit the observed
helium and magnesium spectral lines (see Fig.~\ref{fig:55806}). As a matter of
fact, all the helium lines in the studied spectral range show unusual
line-shapes probably related to the presence of a close companion.



\subsection{HD 178479}

Only two publications are found in SIMBAD for HD~178479, which is
classified B9 V. The strength of the \ion{He}{i} spectral lines
is however much too large for a late B-type star and rather
corresponds to a B3 star (see Fig.~\ref{fig:cal}).

\subsection{HD 179343}

Thin features superimposed on broader spectral lines are detected in the
spectra of HD179343, which is considered as a Be shell star. It is however
interesting to note that it is also known as a "single-entry" binary in the
HIPPARCOS catalogue with $\Delta$V~=~0.481. The shell features could therefore
be an artifact of the secondary component.

\subsection{HD 184279}

HD~184279 is an early B-type star showing numerous variable shell
absorptions superimposed on the photospheric spectrum
\citep{1989A&A...214..285B}, as can be noticed from
Fig.~\ref{fig:eqw} where the equivalent widths of the \ion{He}{i}
4471 and \ion{Mg}{ii} 4481 lines are clearly overestimated. Since
these features are also affecting the helium lines (i.e. our main
temperature and \vsini~criteria), the fitting zones were adapted
to exclude the features as well as possible.


\begin{figure}
\center
\includegraphics[width=7cm,clip=]{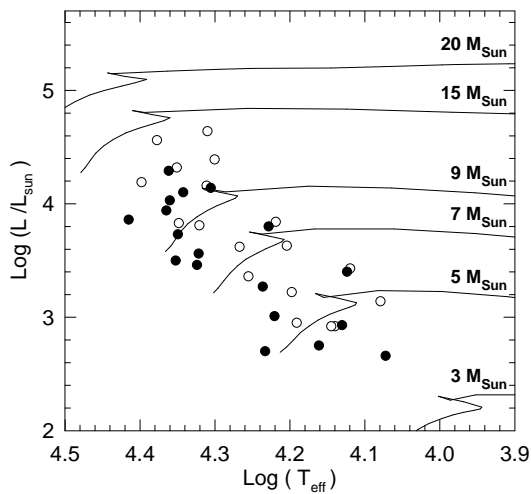}
\caption{Location of Be stars showing the strongest H$\gamma$
emission in the HR diagram with (filled circles) and without
(open circles) accounting for the veiling effects. {\revised
Theoretical evolutionary tracks are taken from
\citet{1992A&AS...96..269S} for a solar-like metallicity.}}
\label{fig:veiling}
\end{figure}

\section{Discussion}

\label{sec:discussion}

\subsection{Effects of veiling correction}

Be stars showing strong H$\gamma$ emission were corrected using
an empirical approach described by \citet{1995A&AS..111..423B}
(see Sect.~\ref{sec:proc:veil}). Including such veiling
corrections in the calculations often produces a lowering of the
observed continuum and, consequently, a spectral line
strengthening. From this procedure, different \teff~and higher
\logg~values are generally obtained, which leads, in the HR
diagram, to pull the location of the Be stars towards the ZAMS
(see Fig.~\ref{fig:veiling}).

\begin{table}
\caption{Average \vsini~{\revised in km~s$^{-1}$} computed
assuming different \omc~towards the Galactic centre and
anticentre. {\revised For each value of \omc, only a certain
number of stars (given in brackets) have a reasonable model (see
Table~\ref{tab:gdcor}) and could be used in the
computation.}\label{tab:vdist}}
\begin{center}
\begin{tabular}{ccc}
\hline\hline\noalign{\smallskip}
 & Anticentre & Centre\\
 \omc & B9--B0 & B9--B0\\
 \hline\noalign{\smallskip}
 0.80  &  221$\pm$65 (27)     &  188$\pm$75 (27)      \\
 0.90  &  236$\pm$75 (29)     &  216$\pm$79 (30)       \\
 0.95  &  251$\pm$84 (31)     &  223$\pm$82 (30)    \\
 0.99  &  260$\pm$92 (32)     &  235$\pm$84 (31)        \\
 \hline
\end{tabular}
\end{center}
\end{table}

\begin{figure*}[ht]
\center
\includegraphics[width=14cm,clip=]{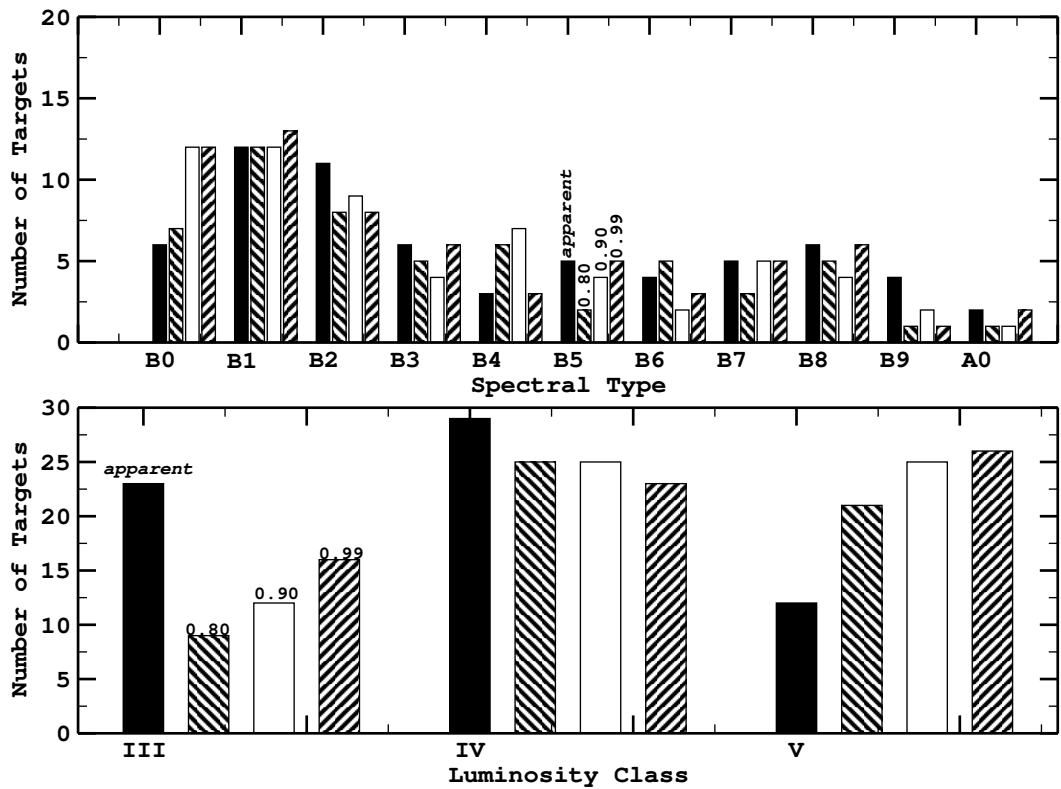}
\caption{\revised Histogram showing the changes in spectral type
and luminosity class distribution when including the effects of
stellar flattening and gravitational darkening. {\it Apparent}
spectral types and luminosity classes are represented with filled
bars, while other bars give the corrected ones adopting \omc=
0.80, 0.90, 0.99. \label{fig:cl}}
\end{figure*}

\subsection{Effects of fast rotation}

In our sample, the effects of fast rotation on effective
temperature, surface gravity and projected rotation velocity are
generally significant when \vsini~$>$ 150~\kps. As expected,
accounting for these effects provides higher values of the $pnrc$
fundamental parameters (e.g. average \vsini~in
Table~\ref{tab:vdist}). {\revised We further note from
Table~\ref{tab:gdcor} that, for \omc $\ge$ 0.8~and if we except
\vsinio, the pnrc stellar parameters generally do not vary
significantly with \omc. This is mainly due to the fact that, at a
fixed \vsini~value, the increase of \omc~leads to smaller
inclinations and, consequently, leads to explore regions of the
stellar surface that are less affected by the flattening of the
star. Since a recent study \citep{effets} showed that Be stars
are found to rotate, on average, at \omc$\sim$ 0.88 when properly
treating fast rotation effects, uncertainties on the actual value
of the angular velocity are not expected to carry too high errors
on the estimate of the $pnrc$ \top~and \gop~parameters.}



{\revised  From Fig.~\ref{fig:cl}, we see that the effects of fast
rotation also appear on the targets spectral type and luminosity
class. As already mentioned by \citet{1994IAUS..162..356B}, the
top of the spectral type distribution of Be stars is centered on
B1 (instead of B2) when gravitational darkening and stellar
flattening are taken into account. It is worth recalling, that
the luminosity class of Be stars are even more sensitive to these
effects, which shift the targets toward lower luminosities when
they are included in the computations. In view of these results,
and keeping in mind that the veiling effect makes this situation
even worse, it seems that luminosity classes are very inaccurate
in characterizing the evolutionary status of Be stars.}


\begin{figure*}[ht]
\begin{center}
\includegraphics[width=14cm,clip=]{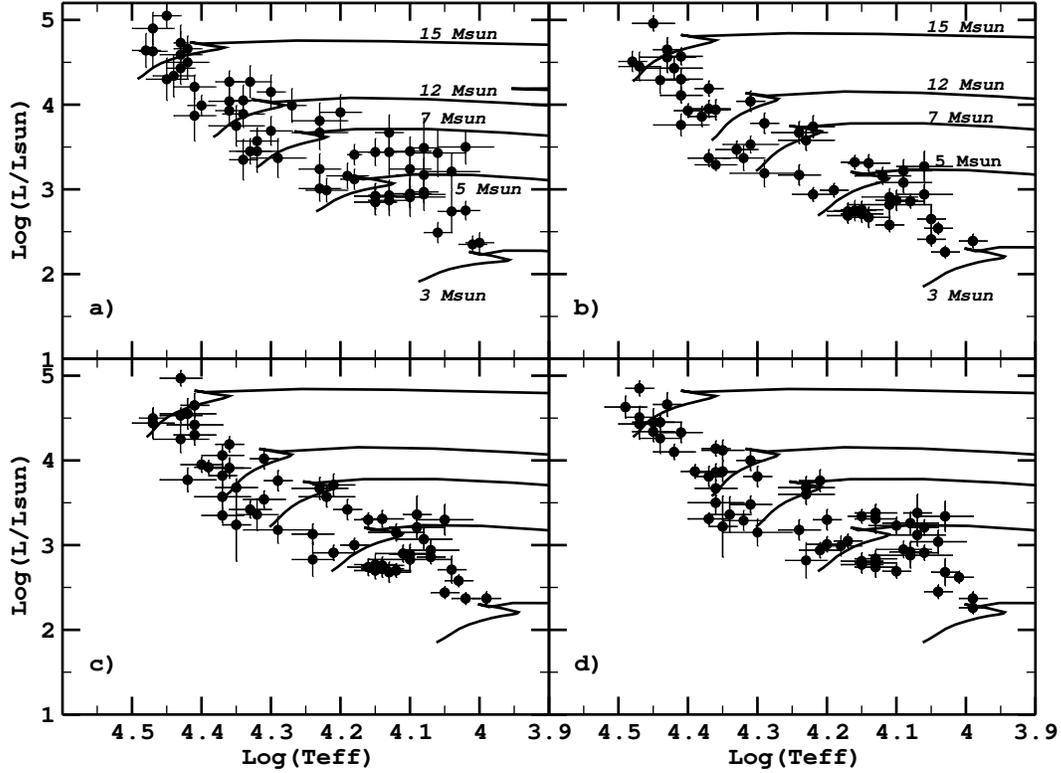}
\caption{\revised Panel a): Location of the Be stars (filled
circles) in the HR diagram adopting the veiling-corrected
apparent stellar parameters of Table~\ref{tab:param}. Theoretical
evolutionary tracks (lines) are taken from
\citet{1992A&AS...96..269S} and adapted following a procedure
given by \citet{zorecevol}. Panels b), c) and d): Location of the
Be stars (filled circles) in the HR diagram accounting for
gravitational darkening effects and assuming \omc~= 0.80, 0.90,
and 0.99, respectively. Theoretical evolution tracks (lines) take
into account the effects of fast rotation as described by
\citet{2000A&A...361..101M}. The effective temperature and the
luminosity reported in the figures are therefore surface-averaged
quantities (Table \ref{tab:jean}).} \label{fig:hres}
\end{center}
\end{figure*}



\begin{table*}
\center \caption{Sample of the $pnrc$ surface-$averaged$
parameters and of the respective interpolated masses
$M/M_{\odot}$, ages $\tau$, and fractional ages $\tau/\tau_{\rm
MS}$. When the projected rotation velocity was greater than the
break-up speed or than the equatorial speed, the Table were left
blank. } \label{tab:jean}
\begin{tabular}{lrrrrrr}
\hline \hline\noalign{\smallskip}
 HD & \teff $^{\rm surf.}$ & \logg $^{\rm surf.}$ & $\log
 L/L_\odot$ $^{\rm surf.}$ & M/M$_\odot$ & age (years) & $\tau / \tau_{\rm
 MS}$\\ \noalign{\smallskip}
 \hline
\multicolumn{7}{|c|}{\omc\ = 0.80}\\
 \hline
 42406 \\
 43264 \\
 43285 & 14500$\pm$1100 & 4.02$\pm$0.17 &  2.69$\pm$ 0.09 &  4.60$\pm$ 0.30 & 6.6E+07$\pm$2.0E+07 &  0.47$\pm$ 0.13 \\
 44783 & 11500$\pm$1100 & 3.33$\pm$0.21 &  2.94$\pm$ 0.12 &  4.30$\pm$ 0.30 & 1.7E+08$\pm$1.5E+07 &  1.02$\pm$ 0.03 \\
 45901 & 26000$\pm$2300 & 3.74$\pm$0.21 &  4.43$\pm$ 0.13 & 13.60$\pm$ 1.30 & 1.2E+07$\pm$2.0E+06 &  0.75$\pm$ 0.10 \\
 46380 & 23000$\pm$1100 & 3.88$\pm$0.14 &  3.94$\pm$ 0.12 & 10.00$\pm$ 0.60 & 1.7E+07$\pm$2.6E+06 &  0.62$\pm$ 0.09 \\
 46484 & 27000$\pm$1000 & 3.64$\pm$0.13 &  4.65$\pm$ 0.14 & 15.70$\pm$ 1.30 & 1.1E+07$\pm$6.0E+05 &  0.80$\pm$ 0.04 \\
 47054 & 12500$\pm$ 800 & 3.58$\pm$0.15 &  2.87$\pm$ 0.12 &  4.60$\pm$ 0.20 & 1.3E+08$\pm$1.1E+07 &  0.91$\pm$ 0.05 \\
 47160 & 11500$\pm$ 600 & 3.74$\pm$0.11 &  2.41$\pm$ 0.08 &  3.60$\pm$ 0.10 & 2.1E+08$\pm$1.7E+07 &  0.79$\pm$ 0.05 \\
47359\\
 49330 & 26000$\pm$1700 & 3.84$\pm$0.17 &  4.30$\pm$ 0.12 & 12.80$\pm$ 1.00 & 1.2E+07$\pm$1.9E+06 &  0.69$\pm$ 0.10 \\
\multicolumn{7}{l}{... {\it A complete version of the table is available at the CDS}}\\
\hline
\end{tabular}
\end{table*}

\subsection{Evolutionary status of Be stars}

The knowledge of whether the Be phenomenon is an innate property
of the stars or {\revised whether} it depends on stellar
evolution is a matter of great interest to better understand the
precise nature of Be stars and its potential link with
asteroseismology. A recent study (Zorec et al. 2005) performed on
a large sample of field Be stars (i.e. 97 Be stars) uniformly
spread over the whole sequence of B type stars showed that the
appearance of the Be phenomenon for the lower mass stars
generally occurs during the second half of the main sequence life
time ($\tau_{\rm MS}$), but that it appears earlier at greater
stellar masses. In order to see if there is such a trend in our
sample of Be stars, we derive their masses, ages and luminosity
using an interpolation procedure developed by Zorec et al. (2005),
which accounts for the changes introduced by fast rotation in the
evolutionary tracks
\citep{2000ApJ...544.1016H,2000A&A...361..101M,2002A&A...390..561M,
2001A&A...373..555M}. Since these modeled evolutionary tracks are
given in terms of surface-$averaged$ effective temperatures and
bolometric luminosities, we transformed the {\it pnrc} parameters
into surface-$averaged$ ones using the angular velocity ratios
$\Omega/\Omega _{\rm c} =$ 0.80, 0.90, and 0.99. The $pnrc$
surface-$averaged$ parameters and the respective interpolated
masses $M/M_{\odot}$, ages $\tau$, and fractional ages
$\tau/\tau_{\rm MS}$ are given in Table \ref{tab:jean}. Fig.
\ref{fig:hres}a shows the HR diagram of the program stars depicted
by the veiling-corrected {\it apparent} $T_{\rm eff}$ and $\log
L/L_{\odot}$ values and evolutionary tracks for non-rotating
stars \citep{1992A&AS...96..269S}. Fig. \ref{fig:hres}b,
\ref{fig:hres}c, and \ref{fig:hres}d show the HR diagram of the
program stars in terms of surface-$averaged$ fundamental
parameters assuming the stars rotate at $\Omega/\Omega_{\rm c} =$
0.80, 0.90, and 0.99 respectively. The corresponding evolutionary
tracks (Fig. \ref{fig:hres}b, \ref{fig:hres}c, and
\ref{fig:hres}d) were calculated for the ZAMS equatorial rotation
velocity $V_o =$ 300 km~s$^{-1}$. Note that $\Omega/\Omega_{\rm
c} =$ 0.90 represents the average angular velocity rate of
galactic field Be stars (Fr\'emat et al. 2005). It must also be
noted that if a star starts its evolution on the ZAMS as a rigid
rotator with a rotation velocity $V_o$, as a consequence of the
initial angular momentum redistribution, the surface velocity
decreases somewhat in a lapse of time ranging from 1 to 2\%
\citep{1999A&A...341..181D} of the stellar $\tau_{\rm MS}$.
Since, on one hand, we do not know the actual initial velocity of
the sample stars on the ZAMS and, on the other hand, dwarf Be
stars have a very flat distribution of their equatorial true
rotational velocities versus the spectral type around $V \sim$ 300
km~s$^{-1}$ {\revised \citep[][ Fig.
1b]{2001A&A...368..912Y,2004IAUS..215...89Z}}, the comparisons
done above with evolutionary tracks calculated for $V_o =$ 300
km~s$^{-1}$ are realistic \citep{2000A&A...361..101M}.\par
\begin{figure}
\center \centerline{\includegraphics[width=8.7cm,clip=]{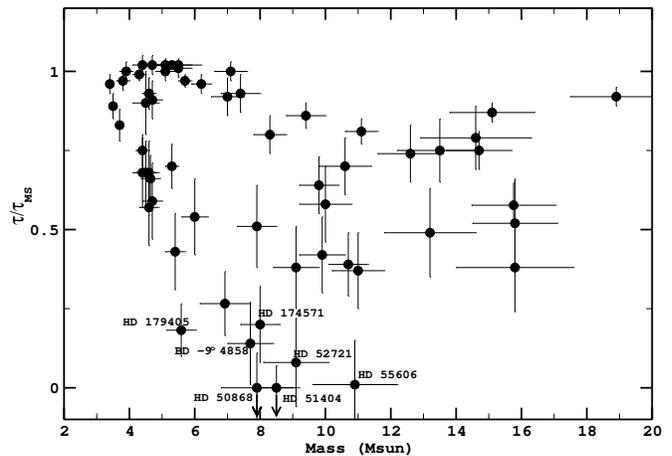}}
\caption[]{$\tau/\tau_{\rm MS}$ ratios of the program Be stars as
a function of stellar mass (see Table~\ref{tab:jean}). The ratios
and masses are computed for \omc~= 0.90, which corresponds to the
expected average angular velocity of Be stars.}
\label{fig:zamstams}
\end{figure}

 Fig.
\ref{fig:zamstams} shows the average $\tau/\tau_{\rm MS}$ ratios
we computed for each star assuming \omc~= 0.90. {\revised Most of
the Be stars are found, on the average, in the second half of the
main sequence life time and we clearly observe a paucity of very
young Be stars at masses lower than 6 $M_{\odot}$. The lack of
low mass Be stars ($M < 7M_{\odot}$) in the first half of the
main sequence was noted in other recent works that deal with
independent stellar samples. Using a magnitude-limited sample (97
Be stars) that mirrors the distribution of all known Be stars in
the Bright Stars Catalogue \citep{BSC2001}, \citet{zorecevol}
obtained a void of low mass Be stars in the first half of the
main sequence evolutionary phase. An equivalent conclusion was
also obtained by \citet{2004serd.book.....Z} from a study of 130
Be stars with visual magnitudes ranging from 7 to 9 mag.} In our
case, probably due to the fact that the sample is smaller and the
stars are not uniformly distributed over the mass range that
defines B-type stars, {\revised the trend is mainly obvious at
masses $< 6 M_{\odot}$. However, since the same result also
repeats in other studies, we think that this effect can be hardly
related to some sampling bias.}


Furthermore, if we exclude the targets with low S/N spectra, 5
stars (HD 50868, HD 51404, HD 52721, HD 174571, BD -9$^o$4858)
are obviously much younger than the rest of the sample
(Fig.~\ref{fig:zamstams}) and might be considered as Herbig Ae/Be
candidates, as is already the case for HD~174571
\citep{2003AJ....126.2971V}.

\subsection{{\revised Be stars as COROT targets}}

{\revised Fields to be observed with COROT have already been
pre-selected based on the parameters of the potential targets.
Only Be stars located in these fields can still be chosen as
secondary targets and would then be observed for about 5
consecutive months. Table~\ref{tab:sec} summarizes these Be stars
that are candidates as secondary targets. Whether all these
fields will be observed with COROT will depend on the length of
the mission; at least five of them will be observed. Moreover,
since COROT tries to sample as well as possible the HR diagram,
not all the secondary Be candidates in a selected field will
eventually be observed with COROT. The choice will be made
according to technical constraints of the satellite and to the
stellar parameters determined in this paper.

Be stars that will not be observed in a long run, either because
they are not located in one of the selected fields (and therefore
are not listed in Table~\ref{tab:sec}) or because they are not
chosen as a secondary target in these fields, can be proposed as
targets for short runs lasting 20-30 days. Again the choice of
these Be short-run targets will depend on the fundamental
parameters determined in this paper, and on their variability. In
particular HD\,168797 seems to be a prime candidate for a
short-run \citep[see][]{gutierrez}.}

\begin{table}
\center
\caption{Be stars close to primary COROT target
candidates, which could be selected as secondary
targets.\label{tab:sec}}
\begin{tabular}{@{}llll@{}}
\hline\hline\noalign{\smallskip}
&\multicolumn{3}{l}{Galactic Centre}\\
Be Secondary & V & Sp. Type & COROT Field \\
\hline\noalign{\smallskip}
HD 171219 & 7.65 & B5 III & HD 171234 \& HD 170580 \\
HD 171219 & 7.65 & B5 III & HD 171834 \\
HD 175869 & 5.56 & B8 III & HD 175726 \\
HD 181231 & 8.58 & B5 IV  & HD 181555 \& HD 180642 \\
\hline\noalign{\smallskip}
&\multicolumn{3}{l}{Galactic Anticentre}\\
Be Secondary & V & Sp. Type & COROT Field \\
\hline\noalign{\smallskip}
HD 43285 & 6.05 & B5 IV & HD 43587 \\
HD 43913 & 7.88 & A0    & HD 43587   \\
HD 45901 & 8.87 & B0.5 IV & HD 46558  \\
HD 46484 & 7.65 & B0.5 IV & HD 46558   \\
HD 47359 & 8.87 & B0 IV & HD 46558  \\
HD 49330 & 8.95 & B0.5 IV & HD 49933 \& HD 49434  \\
HD 49567 & 6.15 & B3 III & HD 49933 \& HD 49434  \\
HD 49585 & 9.13 & B0.5 IV & HD 49933 \& HD 49434 \\
HD 50087 & 9.08 & B8 III & HD 49933 \& HD 49434  \\
HD 50209 & 8.36 & B8 IV &  HD 49933 \& HD 49434  \\
HD 50891 & 8.88 & B0.5 V & HD 52265 \\
HD 51193 & 8.06 & B1.5 IV &  HD 52265   \\
HD 51404 & 9.30 & B1.5 V &  HD 52265 \\
HD 51452 & 8.08 & B0 IV &  HD 52265 \\
\hline
\end{tabular}
\end{table}

\section{Conclusions}

In this paper, we determined the fundamental parameters (spectral
type, effective temperature, surface gravity and projected
rotation velocity) for the Be stars that can be observed in
COROT's seismology fields. To this end, a careful and detailed
modeling of the stellar spectra was applied, accounting for NLTE
effects as well as for the effects due to fast rotation (stellar
flattening and gravitational darkening). {\revised On the
average, the uncertainties we obtained on the stellar parameters
are of the order of 10 \% on \vsini, 7 \% on \teff, and 3.3 \% on
$\log L/L_{\odot}$. Our results are therefore very well suited
for the target selection of COROT Be stars and, further, will
also be very useful when analyzing the huge amount of data that
COROT will provide.}

{\revised A study of the fast rotation effects, as well as of
those related to circumstellar veiling, on the spectral type and
luminosity class distribution shows that the luminosity class of
Be stars is generally inaccurate in characterizing their
evolutionary status.} Evolutionary tracks for fast rotators were
further used to derive stellar masses and ages in order to
discuss the appearance of the Be phenomenon with respect to the
evolutionary status of the stars. {\revised Most of the Be stars
of our sample are found in the second half of the main sequence
life time. There is further an obvious lack of very young Be
stars at masses lower than 6 $M_{\odot}$ which reproduces previous
observations made by \citet{zorecevol} and
\citet{2004serd.book.....Z}.}


Finally, we give a preliminary list of the Be stars that could be
chosen as secondary targets for the COROT mission. The
fundamental stellar parameters we derived will be further used to
carry on the secondary target selection procedure, as well as to
determine which object could be proposed for short runs.

\begin{acknowledgements}
We wish to acknowledge the referee, D. Baade, for his suggestions
which greatly improved the readability of the paper. We thank
J.Guti\'errez--Soto for his remarks. We are grateful to the
ground-based asteroseismology working group of COROT and all the
contributors of GAUDI. This research has made use of the SIMBAD
database maintained at CDS, Strasbourg, France. YF thanks Dr.
P.Lampens for hosting him at the Royal Observatory of Belgium.
JRM acknowledges continuous financial support of the CNPQ
Brazilian Agency.
\end{acknowledgements}

\bibliographystyle{aa}
\bibliography{cocor}

\newpage

\onecolumn \center

 {\Huge ONLINE DATA}


\newpage

\setcounter{table}{0}

\begin{table}[t]
\caption[]{Spectra used in this study. {\revised For the ELODIE
spectra, the signal to noise ratio (S/N) was provided by the
INTERTACOS (OHP) reduction pipeline, while for the other data it
was computed with IRAF by selecting some parts of the continuum in
the studied spectral region.} '$^1$' indicates that the
spectrum is available in GAUDI. } 
\begin{center}
\begin{tabular}{llllr}
\hline \hline\noalign{\smallskip} HD & Obs. date & T$_{\rm exp}$ (s) & Instrument & S/N \\
\hline\noalign{\smallskip}
42406     & 2004-02-05 & 900      & FEROS &249\\  
43264     & 2001-11-27 & 3300  & ELODIE$^1$ &83\\
43285     & 2001-12-21 & 1800  & ELODIE$^1$ &123\\
44783     & 2000-12-18 & 1500  & ELODIE$^1$ &121\\
45901     & 2004-01-03 & 2700  & AURELIE 4280\_G3&120\\
46380     & 2001-12-22 & 3600  & ELODIE$^1$ &53\\
46484     & 2003-01-26 & 3600  & ELODIE$^1$ &107\\
47054     & 2002-01-28 & 300   & FEROS$^1$ &145\\
47160     & 2002-01-28 & 300   & FEROS$^1$ &150\\
47359     & 2004-01-03 & 2400  & AURELIE 4280\_G3&140\\
49330     & 2001-12-22 & 3600  & ELODIE$^1$ &54\\
49567     & 2002-04-01 & 1500  & ELODIE$^1$ &132\\
49585     & 2004-01-07 & 3600  & AURELIE 4485\_G2&140\\
49787     & 2003-01-18 & 220   & FEROS$^1$ &126\\
50083     & 2003-01-17 & 1200  & ELODIE$^1$ &96\\
50138     & 2003-01-18 & 110   & FEROS$^1$ &144\\
50209     & 2001-12-24 & 5150  & ELODIE$^1$ &46\\
50581     & 2003-01-27 & 3600  & ELODIE$^1$ &73\\
50696     & 2001-12-22 & 7200  & ELODIE$^1$ &48\\
50868     & 2001-10-09 & 500   & FEROS$^1$ &160\\
50891     & 2005-01-05 & 4500      & AURELIE &250\\
51193     & 2004-01-03 & 900   & AURELIE 4280\_G3&130\\
51404     & 2004-01-04 & 3600  & AURELIE 4280\_G3&150\\
51452     & 2004-01-03 & 900   & AURELIE 4280\_G3&126\\
51506     & 2003-01-24 & 3600  & ELODIE$^1$ &89\\
52721     & 2004-02-05 & 300      & FEROS &260\\ 
53085     & 2003-01-15 & 180   & FEROS$^1$ &115\\
53667     & 2003-01-18 & 300   & FEROS$^1$ &105\\
54464     & 2004-02-05 & 1800   & FEROS&317\\
55135     & 2004-02-05 & 300   & FEROS&250\\
55606     & 2001-12-22 & 5400  & ELODIE$^1$ &40\\
55806     & 2004-02-05 & 1500   & FEROS &250\\ 
57539     & 2003-01-17 & 130   & FEROS$^1$ &72\\
166917    & 2002-08-14 & 720   & ELODIE$^1$ &90\\
168797    & 1999-07-25 & 900   & ELODIE$^1$ &171\\
170009    & 2000-06-07 & 2400  & ELODIE$^1$ &141\\
170714    & 2001-07-06 & 450   & FEROS$^1$ &98\\
171219    & 2002-07-07 & 300   & FEROS$^1$ &130\\ 
173219    & 2001-07-06 & 450   & FEROS$^1$ &85\\
173371    & 2001-07-06 & 300   & FEROS$^1$ &115\\
173530    & 2003-07-18 & 600   & OPD CASS &30\\ 
173637    & 2003-07-18 & 1200  & OPD CASS &50\\
173817    & 2003-07-18 & 600   & OPD CASS &70\\
174513    & 2004-07-05 & 2740  & AURELIE 4481\_G3&170\\
174571    & 2004-07-12 & 3600  & AURELIE 4481\_G3&180\\
          & 2004-07-15 & 3600  & AURELIE 4100\_G3&200\\
\hline
\end{tabular}
\end{center}
\end{table}

\setcounter{table}{0}
\begin{table}[t]
\begin{center}
\begin{tabular}{llllr}
\hline \hline\noalign{\smallskip}
HD & Obs. date & T$_{\rm exp}$ (s) & Instrument & S/N\\
\hline\noalign{\smallskip}
174705    & 2004-07-05 & 2400  & AURELIE 4481\_G3&140\\
          & 2004-07-15 & 3600  & AURELIE 4100\_G3&140\\
174886    & 2002-07-07 & 300   & FEROS$^1$ &80\\
175869    & 2001-07-07 & 150   & FEROS$^1$ &90\\
176159    & 2004-07-20 & 2500  & AURELIE 4481\_G3&190\\
176630    & 2001-07-07 & 1800  & ELODIE$^1$ &94\\
          & 2001-07-06 & 600   & FEROS$^1$ &100\\ 
178479    & 2004-07-21 & 3600  & AURELIE 4481\_G3&180\\
179343    & 2003-07-18 & 400   & OPD CASS &120\\
179405    & 2004-07-04 &       & AURELIE 4481\_G3 & \\  
180126    & 2004-07-04 & 1000  & AURELIE 4481\_G3&320\\
          & 2004-07-13 & 2700  & AURELIE 4100\_G3&170\\
181231    & 2003-07-18 & 1000  & OPD CASS &60\\
181308    & 2003-07-18 & 700   & OPD CASS &70\\ 
181367    & 2003-07-18 & 1000  & OPD CASS &50\\
181709    & 2004-07-20 & 2000  & AURELIE 4481\_G3&180\\
181803    & 2004-07-19 & 3600  & AURELIE 4481\_G3&125\\ %
184279    & 2002-07-06 & 300   & FEROS$^1$ &40\\ 
          & 2003-07-18 & 500   & OPD CASS &100\\
184767    & 2004-07-18 & 1000  & AURELIE 4481\_G3&110\\
194244    & 1999-12-20 & 3600  & ELODIE$^1$ &112\\ 
230579    & 2004-07-05 & 3600  & AURELIE 4481\_G3&130\\
BD-09$^{o}$4858 & 2004-07-20 & 2400  & AURELIE 4481\_G3&180\\
\hline
\end{tabular}
\end{center}
\end{table}

\setcounter{table}{3}
\begin{table*}[t]
\caption[]{\revised Veiling corrected apparent stellar parameters.
ID numbers, SIMBAD V magnitudes and spectral types are given for
each target in cols. 1, 2, 3 and 4. The derived stellar parameters
(effective temperature, surface gravity, and \vsini) are gathered
in cols. 5, 6 and 7. Their accuracy is estimated by scanning the
solutions space while adopting different initial values for the
parameters. Spectral types (col. 8) are derived from the apparent
stellar parameters combined to the \teff~and \logg~calibrations
proposed by \citet{1994AJ....107..742G} and by
\citet{1986serd.book.....Z}. Cols. 9 and 10 list the equivalent
width of the H$\gamma$ emission components as well as the
estimates of the veiling correction, respectively. The error bars
on the equivalent widths are generally of the order of 15\% and
are a product of the fitting process. Errors on the veiling
parameter, r, are estimated by accounting for the accuracy on
W$_{\lambda}^{\rm{H}\gamma}$ and by assuming a 95\% confidence
interval on the reference data of \citet[Fig.
9a,][]{1995A&AS..111..423B}. Previous determinations of the
stellar parameters are given in col.11 and are taken from: (1)
\citet{2002A&A...393..897R}; (2) \citet{1995A&AS..110..553S}; (3)
\citet{2002ApJ...573..359A}; (4) \citet{2001A&A...368..912Y}; (5)
\citet{2004AJ....127.1176L}; (6) \citet{2001A&A...378..861C}; (7)
\citet{2002MNRAS.333....9L}; (8) \citet{1989A&AS...80...73G}; (9)
\citet{1986serd.book.....Z}; (10) \citet{1996PASP..108..833H};
(11) \citet{2001A&A...368..160C}; (12)
\citet{1977ApJ...213..438C}; (13) \citet{1985MNRAS.214..327T};
(14) \citet{1982ApJS...50...55S}.}
\begin{center}
\begin{tabular}{@{ }r@{ }r@{ }cl@{ }l@{ }cl@{ }l@{ }rlrrl @{ }}
\hline\hline
\noalign{\smallskip} \multicolumn{2}{c}{ID} && \multicolumn{2}{c}{SIMBAD} && \multicolumn{4}{c}{This work}\\
\cline{1-2}\cline{4-5}\cline{7-10}\noalign{\smallskip}
HD/BD & HIP/BD/    && V     & Sp. Type      && T$_{\rm eff}$ & \logg & V sin {\it i} & Sp. Type   & W$_{\lambda}^{\rm{H}\gamma}$     &   r  &  Notes\\
      &  MWC       &&       &               && (K)           &    (c.g.s.) & (km s$^{\rm{-1}}$)  &     & (m\AA) &     \\
          \hline\noalign{\smallskip}

42406 & 29298      && 8.01  & B9        && 15400$\pm$1000 & 3.72$\pm$0.10  &  300$\pm$25 & B4 IV       & 0.30$\pm$0.05 & -- \\
43264 & 29719      && 7.51  & B9        && 10500$\pm${\revised 1000} & 2.76$\pm$0.10 & 288$\pm$10   & B9 III & 0.22$\pm$0.03 & --    \\
43285 & 29728      && 6.05  & B6Ve      && 14000$\pm$1000 & 3.78$\pm$0.10 & 260$\pm$20  & B5 IV& 0.21$\pm$0.03 & -- & \vsini~= 235 \kps~(3)     \\ 
      &            &&       &           && && && &  & \vsini~= 260 $\pm$ 12 \kps~(4)\\
      &            &&       &           && && && &  & \teff~= 16600 $\pm$ 600 \kps~(5)\\
      &            &&       &           &&&& && &  & \logg~= 4.0 $\pm$ 0.1 (5)\\
      &            &&       &           &&&& && &  & \vsini~= 237 $\pm$ 11 \kps~(5)\\
44783 & 30448      && 6.225 & B8Vn      && 11000$\pm$1000 & 3.05$\pm$0.15 & 226$\pm$50 & B9 III & 0.10$\pm$0.02 & --  & \vsini~=230 \kps~(3)    \\ 
45901 & 30992      && 8.87  & B2Ve      && 26500$\pm$2000 & 3.73$\pm$0.15 & 164$\pm$15 & B0.5 IV & 0.82$\pm$0.12 & 0.17$\pm$0.08 \\
46380 & 31199      && 8.05  & B2Vne     && 22000$\pm$1000 & 3.70$\pm$0.15 & 300$\pm$15  & B1.5 IV & 0.91$\pm$0.14 & 0.20$\pm$0.08 & \teff~=21200 $\pm$ 650 K (5)\\
      &            &&       &           &&&& & & & & \logg~=3.5 $\pm$ 0.1 (5)\\
      &            &&       &           && && & & & & \vsini~= 293 $\pm$ 3 \kps~(5)\\ 
      &            &&       &           &&&& & & & & \vsini~= 262 \kps~(4)\\
46484 & 31305      && 7.65  & B1V       && 27000$\pm$900 & 3.60$\pm$0.15 & 120$\pm$20 & B0.5 IV & 0.15$\pm$0.02 & --      \\ 
47054 & 31583      && 5.57  & B8Ve      && 12000$\pm$700 & 3.40$\pm$0.15 & 222$\pm$6 & B7 III & 0.30$\pm$0.05 & -- & \vsini~= 220 \kps~(3)          \\
      &            &&       &           &&&& & & & & \vsini~= 226 \kps~(4)\\
      &            &&       &           &&&& & & & & \teff~= 11995 K (6)\\
      &            &&       &           &&&& & & & & \logg~= 3.9 (6)\\
      &            &&       &           &&&& & & & & \vsini~= 230 \kps~(6)\\
47160 & 31629      && 7.104 & B9        && 11500$\pm$500 & 3.70$\pm$0.10 & 149$\pm$5 & B8 IV & 0.07$\pm$0.01 & --    \\ 
47359 & +05\d1340  && 8.87  & B0.5Vpe...    && 29500$\pm$2000    & 3.65$\pm$0.15     &   443$\pm$40 & B0 IV & 0.41$\pm$0.06 & --      \\
49330 & 32586      && 8.95  & B0:nnpe       && 27000$\pm$1500 & 3.82$\pm$0.15 & 270$\pm$50 & B0.5 IV  & 0.25$\pm$0.04 & -- & \teff~= 27200 $\pm$ 600 K (5)     \\ 
      &            &&       &           &&&& & && &  \logg~= 4.0 $\pm$ 0.1 (5)\\
      &            &&       &           &&&& & && &  \vsini~= 200 $\pm$ 10 \kps~(5)\\
49567 & 32682      && 6.146 & B3II-III      && 17000$\pm$1500 & 3.48$\pm$0.12 & 85$\pm$10 & B3 III  & 0.23$\pm$0.03 & -- & \vsini~= 75 \kps~(3)    \\ 
      &            &&       &           &&&& & && &  \teff~= 16600 $\pm$ 500 K (7)\\
      &            &&       &           &&&& & && &  \logg~= 3.31 $\pm$ 0.19 (7)\\
      &            &&       &           &&&& & && &  \teff~= 17300 $\pm$ 200 K (8)\\
      &            &&       &           &&&& & && &  \teff~= 16157 K (9)\\
      &            &&       &           &&&& & && &  \logg~= 2.95 -- 3.10 (9)\\
49585 & +00\d1624  && 9.13  & B0.5:V:nn     && 25500$\pm$1500 & 3.87$\pm$0.1 & 310$\pm$30 & B0.5 IV & 0.10$\pm$0.02 & -- \\
49787 & 32766      && 7.54  & B1V:pe        && 25000$\pm$1000 & 4.01$\pm$0.10 & 160$\pm$15 & B1 V & 0.12$\pm$0.02 & -- &  \vsini~= 186 $\pm$ 2 \kps~(4)     \\ 
      &            &&       &           &&&& & && &  \vsini~= 180 $\pm$ 2 \kps~(10)\\
50083 & 32947      && 6.92  & B2Ve      && 20000$\pm$1000     &  3.43$\pm$0.15    &  181$\pm$20 & B2 III       & 1.33$\pm$0.20 & 0.30$\pm$0.11 &  \vsini~= 178 $\pm$ 2 \kps~(4)  \\
      &            &&       &           &&&& & && &  \vsini~= 170 $\pm$ 2 \kps~(10)\\
50138 & 32923      && 6.583 & B9        && 12500$\pm$1000 & 3.25$\pm$0.15 & 60$\pm$10 & B7 III  & -- & -- &  \teff~= 13279 $\pm$ 100 K (11)      \\ 
      &            &&       &           &&&& & && &  \logg~= 3.4 (11)\\
50209 & 32977      && 8.36  & B9Ve      && 12500$\pm$1500 & 3.50$\pm$0.10 & 200$\pm$30 & B8 IV  & 0.62$\pm$0.09 & 0.12$\pm$0.07 &  \teff~=12500 $\pm$ 450 K (5)    \\
      &            &&       &           &&&& & && &  \logg~= 4.00 $\pm$ 0.15 (5)\\
      &            &&       &           &&&& & && &  \vsini~= 173 $\pm$ 15 \kps~(5)\\
50581 & 33167      && 7.54  & A0        && 10250$\pm$500 & 3.60$\pm$0.08 & 240$\pm$25 & A0 IV  & 0.30$\pm$0.05 & --    \\ 
50696 & +00\d1691  && 8.87  & B1:V:nne      && 21300$\pm$1500 & 3.45$\pm$0.15 & 350$\pm$30 & B1.5 III  & 0.28$\pm$0.04 & -- &  \teff~= 21700 $\pm$ 600 K (5)  \\ 
      &            && & &&& &           &&& &  \logg~= 3.50 $\pm$ 0.15 (5)\\
      &            && & &&& &           &&& &  \vsini~= 281 $\pm$ 18 (5)\\
50868 & 33267      && 7.85  & B2Vne     && 22000$\pm$1000 & 4.27$\pm$0.15 & 267$\pm$20 & B1.5 V   & 0.02$\pm$0.01 & -- \\
50891 & -03\d1643  && 8.88  & B0:pe     &&  27500$\pm$2000    & 3.93$\pm$0.20     &  220$\pm$20    & B0.5 V   & -- & --  \\
\hline
\end{tabular}
\end{center}
\end{table*}

\setcounter{table}{3}
\begin{table*}[t]

\begin{center}
\begin{tabular}{@{ }r@{ }r@{ }cl@{ }l@{ }cl@{ }l@{ }rlrrl @{ }}
\hline\hline
\noalign{\smallskip} \multicolumn{2}{c}{ID} && \multicolumn{2}{c}{SIMBAD} && \multicolumn{4}{c}{This work}\\
\cline{1-2}\cline{4-5}\cline{7-10}\noalign{\smallskip}
HD/BD   & HIP/BD/   && V     & Sp. Type      && T$_{\rm eff}$ & \logg & V sin {\it i} & Sp. Type   & W$_{\lambda}^{\rm{H}\gamma}$     &   r  &  Notes\\
        &  MWC      &&       &               && (K)           &    (c.g.s.) & (km s$^{\rm{-1}}$)  &     & (m\AA) &     \\
          \hline\noalign{\smallskip}
51193 & 33361      && 8.06  & B1V:nn        &&  23000$\pm$1000 & 3.62$\pm$0.1     & 215$\pm$25 & B1.5 IV & 1.35$\pm$0.20  & 0.23$\pm$0.11     \\
51404 & -06\d1840  && 9.30  & B9        &&  21500$\pm$1500 & 4.15$\pm$0.1     & 335$\pm$10 & B1.5 V  & 0.56$\pm$0.08 & 0.10$\pm$0.07         \\
51452 & -04\d1745  && 8.08  & B0:III:nn     &&  30000$\pm$1500    & 3.88$\pm$0.15     &  298$\pm$20 & B0 IV & 0.23$\pm$0.03 &  --       \\
51506   & 33509     && 7.68  & B5        && 17000$\pm$1000 & 3.84$\pm$0.15 & 172$\pm$20 & B2.5 IV & 0.84$\pm$0.13 & 0.20$\pm$0.08 \\ 
52721   & 33868     && 6.58  & B2Vne         && 22500$\pm$2000      &  3.99$\pm$0.20    &   352$\pm$40 & B1.5 V & 0.59$\pm$0.09 & 0.05$\pm$0.07  & \vsini~= 456 $\pm$ 2 \kps (10)      \\
        &           &&       &           &&             &           &          &   &   &       & \vsini~= 243 $\pm$ 93 \kps (4)      \\
53085   & 34032     && 7.20  & B8        && 15000$\pm$700 & 3.70$\pm$0.10 & 203$\pm$25 & B4 IV & 0.37$\pm$0.06 & --   \\
53667   & -08\d1734 && 7.76  & B0.5III       && 28000$\pm$1500 & 3.45$\pm$0.10 & 110$\pm$10 &B0 III & 0.10$\pm$0.02  & -- & \vsini~= 85 \kps~(12)  \\ 
54464   & -03\d1762 && 8.40  & B2:V:pe       && 17000$\pm$1500 & 3.38$\pm$0.15 & 162$\pm$15 &B2.5 III & 0.64$\pm$0.10  & 0.05$\pm$0.07 &         \\
55135   & 34719     && 7.32  & B4Vne         && 17000$\pm$1000 & 4.34$\pm$0.20 & 244$\pm$20 &B2.5 V & 1.20$\pm$0.18  & 0.35$\pm$0.10 &  \vsini~= 258 $\pm$ 25 \kps~(4)\\
        &           &&       &           && & & &  & && \vsini~= 270 $\pm$ 25 \kps~(10)\\
55606   & -01\d1603 && 9.04  & B1:V:nnpe     && 26000$\pm$2000 & 4.20$\pm$0.20    & 350$\pm$60 & B0.5 V& 1.42$\pm$0.21  & 0.35$\pm$0.12 & \teff~= 28700 $\pm$ 550 K (5)   \\
        &           &&       &           && & & &  & && \logg~= 4.1 $\pm$ 0.1 (5)\\
        &           &&       &           && & & &  & && \vsini~= 335 $\pm$ 20 \kps~(5)\\
55806   & 35021     && 9.10  & B9        && 12500$\pm$1000     &  3.47$\pm$0.10 &  202$\pm$15  & B7 III  & -- & -- & SB2 (Sect. \ref{sec:55806})  \\
57539   & 35669     && 6.581 & B5III         && 14000$\pm$1000 & 3.30$\pm$0.10      & 141$\pm$10 & B5 III & 0.31$\pm$0.05 & --        \\
166917  & 89242     && 6.686 & B9        && 12000$\pm$1000 & 3.23$\pm$0.15 & 165$\pm$10 & B8 III & 0.10$\pm$0.02 & --  & \vsini~= 170 $\pm$ 7 \kps~(10)  \\  
168797  & 89977     && 6.147 & B3Ve      && 18500$\pm$1000 & 3.40$\pm$0.15 & 264$\pm$20 & B2.5 III & 0.09$\pm$0.01 & --  & \teff~= 21000 K (13) \\
        &           &&       &           &&& &&&   & & \vsini~= 260 \kps~(3)\\
        &           &&       &           &&& &&&   & & \vsini~= 251 $\pm$ 6 \kps~(4)\\
170009  & 90428     && 8     & B8        && 11000$\pm$500 & 3.40$\pm$0.20 & 180$\pm$30 & B9 III   & 0.40$\pm$0.06 & --      \\ 
170714  & 90768     && 7.38  & B1Vne         && 23000$\pm$1200 & 3.89$\pm$0.15 & 270$\pm$15 & B1.5 IV &  0.88$\pm$0.13  & 0.15$\pm$0.08   \\
171219  & 90958     && 7.65  & B8        && 13500$\pm$700 & 3.07$\pm$0.15 & 300$\pm$25 & B5 III & -0.20$\pm$0.03 & --  & \teff~= 13500 $\pm$ 500 K (5)  \\
        &           &&       &           &&&   & &&& & \logg~= 3.80 $\pm$ 0.15 (5)\\
        &           &&       &           &&&   & &&& & \vsini~= 190 $\pm$ 25 \kps~(5)\\
173219  & 91946     && 7.88  & B1:V:npe      && 27000$\pm$2000 & 3.70$\pm$0.20 & 61$\pm$10 & B0.5 IV & 0.80$\pm$0.12 & 0.15$\pm$0.08 \\
173371  & 91987     && 6.885 & B9III         && 12500$\pm$500 & 3.50$\pm$0.10 & 295$\pm$10 & B7 IV & 0.29$\pm$0.04 & --  & \vsini~= 291 $\pm$ 4 \kps~(4) \\
173530  & +04\d3870 && 8.87  & B9        && 12500$\pm$1000 & 3.10$\pm$0.10 & 250$\pm$15 & B7 III & 0.28$\pm$0.04 & --   \\
173637  & 92128     && 9.29  & B1IV      && 26500$\pm$1500    & 3.61$\pm$0.10     & 197$\pm$20 & B1 IV  & 1.61$\pm$0.24 & 0.40$\pm$0.14  & \vsini~= 98 $\pm$ 10 \kps~(4)   \\
173817  & 92180     && 8.65  & B8        && 13500$\pm$1000 & 3.70$\pm$0.15 & 270$\pm$15 & B6 IV & -0.15$\pm$0.02 & --    \\
174513  & 92510     && 8.70  & B1V:npe       && 23000$\pm$1500 & 3.80$\pm$0.10     & 251$\pm$30 & B1.5 IV & 0.91$\pm$0.14 & 0.20$\pm$0.08  & 27500 $\pm$ 600 K (5)      \\
        &           &&       &           && &&& &  & & \logg~= 3.5 $\pm$ 0.1~(5)\\
        &           &&       &           && &&& &  & & \vsini~= 180 $\pm$ 15 \kps~(5)\\
        &           &&       &           && &&& &  & & \vsini~= 202 $\pm$ 5 \kps~(4)\\
        &           &&       &           && &&& &  & & \vsini~= 200 \kps~(10)\\
174571  & 92477     && 8.89  & B3V:pe        && 21000$\pm$1500 & 4.00$\pm$0.10     & 240$\pm$15 & B1.5 V & 0.83$\pm$0.12 & 0.15$\pm$0.08    & \vsini~=294 $\pm$ 19 \kps~(4)     \\
174705  & -11\d4786 && 8.34  & B2Vne         && 22000$\pm$1500     & 3.83$\pm$0.10  &   331$\pm$25 & B1.5 IV & 0.05$\pm$0.01 & --    \\
174886  & 92694     && 7.77  & B8        && 15000$\pm$500 & 3.45$\pm$0.10 & 69$\pm$5 & B4 III & 0.35$\pm$0.05 & -- \\
175869  & 93051     && 5.563 & B9IIIpe...    && 12000$\pm$500 & 3.38$\pm$0.10 & 167$\pm$10 & B8 III & 0.15$\pm$0.02 & --   & \vsini~= 140 \kps~(3)\\
        &           &&       &           && &&& &  & & \vsini~= 120 \kps~(4)\\
        &           &&       &           && &&& &  & & \teff~= 11066 K (6)\\
        &           &&       &           && &&& &  & & \logg~= 3.48 (6)\\
        &           &&       &           && &&& &  & & \vsini~= 175 \kps~(6)\\
176159  & 93215     && 8.98  & B9        && 14000$\pm$1000 &  3.73$\pm$0.10    &  227$\pm$15 & B5 IV & 0.39$\pm$0.06 & --    & \vsini~=218 $\pm$ 1 \kps~(4)     \\
        &           &&       &           && &&&&   & & \vsini~= 220 $\pm$ 1 \kps~(10)\\
176630  & 93411     && 7.70  & B4IV      && 16000$\pm$1000 & 3.20$\pm$0.15 & 175$\pm$20 & B3 III & 0.16$\pm$0.02 & --  \\ 
178479  & 94011     && 8.92  & B9V       && 16500$\pm$1000 & 3.99$\pm$0.10      &  99$\pm$5 & B3 V   &  0.87$\pm$0.13 & 0.15     \\
179343  & 94331     && 6.945 & B9        && 11500$\pm$1000 & 2.97$\pm$0.20 & 148$\pm$15 & B8 III   & 0.25$\pm$0.04 & --   & \vsini~= 261 $\pm$ 36 \kps~(4) \\
179405  & 94384     && 9.12  & B5        && 19500$\pm$2000 & 4.01$\pm$0.15     & 231$\pm$20 & B2 V & -- & --     \\
180126  & 94596     && 7.99  & B3p       && 20000$\pm$1500      &  3.80$\pm$0.10    &  243$\pm$20 & B2 IV & -- & --     \\
181231  & 94988     && 8.58  & B9V       && 14000$\pm$1000 & 3.73$\pm$0.10 & 250$\pm$30 & B5 IV  & -- & --     \\
181308  & -01\d3711 && 8.70  & B8        && 14000$\pm$1000 & 3.78$\pm$0.10 & 246$\pm$15 & B5 IV & 1.02$\pm$0.15 & 0.23$\pm$0.09     \\
181367  & +02\d3852 && 9.36  & B8        && 13500$\pm$1000 & 3.65$\pm$0.15 & 279$\pm$30 & B6 IV & 0.55$\pm$0.08 & 0.05$\pm$0.07      \\
181709  & 95133     && 8.79  & B8        && 13500$\pm$1000      &  3.24$\pm$0.20  & 249$\pm$10 & B6 III & 0.60$\pm$0.09 & 0.05$\pm$0.07      \\
181803  & 95152     && 9.1   & B9        && 12000$\pm$500      & 3.$\pm$0.25    & 185$\pm$50 & B7 III  & -- & --        \\  
\hline
\end{tabular}
\end{center}
\end{table*}

\setcounter{table}{3}
\begin{table*}[t]

\begin{center}
\begin{tabular}{@{ }r@{ }r@{ }cl@{ }l@{ }cl@{ }l@{ }rlrrl @{ }}
\hline\hline
\noalign{\smallskip} \multicolumn{2}{c}{ID} && \multicolumn{2}{c}{SIMBAD} && \multicolumn{4}{c}{This work}\\
\cline{1-2}\cline{4-5}\cline{7-10}\noalign{\smallskip}
HD/BD     & HIP/BD/   && V     & Sp. Type  && T$_{\rm eff}$ & \logg & V sin {\it i} & Sp. Type   & W$_{\lambda}^{\rm{H}\gamma}$     &   r  &  Notes\\
          &  MWC      &&       &           && (K)           &    (c.g.s.) & (km s$^{\rm{-1}}$)  &     & (m\AA) &     \\
          \hline\noalign{\smallskip}
184279    & 96196     && 6.98  & B0.5IV    && 28000$\pm$2000    &  4.00$\pm$0.20    &  230$\pm$30  & B0 V    & -- & --  & \teff~=30400 $\pm$ 600 K (5)  \\
      &           &&       &       && &&&&   & & \logg~= 3.9 $\pm$ 0.1 (5)\\
      &           &&       &       && &&&&   & & \vsini~= 200 $\pm$ 22 \kps~(5)\\
      &           &&       &       && &&&&   & & \vsini~= 212 $\pm$ 9 \kps~(4)\\
      &           &&       &       && &&&&   & & \teff~= 30408 K (6)\\
      &           &&       &       && &&& &  & & \logg~= 3.92 (6)\\
      &           &&       &       && &&& &  & & \vsini~= 195 $\pm$ 20 (6)\\
184767    & 96403     && 7.179 & A2    && 10000$\pm$500 & 3.52$\pm$0.10 & 44$\pm$15 & A0 III  & 0.15$\pm$0.02 & --\\ 
194244    & 100664    && 6.144 & B9V       && 10500$\pm$500 & 3.31$\pm$0.08 & 233$\pm$15 & B9 III & 0.25$\pm$0.04 & -- & \vsini~= 222 \kps~(1)\\ 
      &           &&       &       && &&& & & & \vsini~= 175 \kps~(3)\\
      &           &&       &       && &&& & & & \vsini~= 221 $\pm$ 14 \kps~(4)\\
      &           &&       &       && &&& & & & \vsini~= 220 \kps~(14)\\
230579    & +10\d3774 && 9.10  & B1.5:IV:ne&& 29500$\pm$1500      &  3.85$\pm$0.15    &   330$\pm$40 & B1 IV &  0.45$\pm$0.07 & --      \\
-09\d4858 & MWC964    && 8.84  &  B    && 21000$\pm$1500      & 4.10$\pm$0.20     & 108$\pm$20 & B1.5 V & 1.33$\pm$0.20 & 0.35$\pm$0.11   \\
\hline
\end{tabular}
\end{center}
\end{table*}

\newpage

\setcounter{table}{4}

\begin{table*}
\center \caption{Fundamental parameters corrected for the effects
of fast rotation at different \omc~ratios. {\revised Error bars
on the parameters are of the same order than in
Table~\ref{tab:param}. When the projected rotation velocity was
greater than the break-up speed or than the equatorial speed, the
cells were left blank.}}
\begin{tabular}{@{\ }ll@{\ \ }l@{\ \ }l@{\ }l@{\ \ }cl@{\ \ }l@{\ \ }l@{\ }l@{\ \ }cl@{\ \ }l@{\ \ }l@{\ }l@{\ \ }cl@{\ \ }l@{\ \ }l@{\ }l@{\ }}
\hline \hline
   & \multicolumn{4}{c}{\omc~= 0.80} && \multicolumn{4}{c}{\omc~= 0.90} && \multicolumn{4}{c}{\omc~= 0.95} && \multicolumn{4}{c}{\omc~= 0.99}\\
\cline{2-5}\cline{7-10}\cline{12-15}\cline{17-20}\\
HD &  \top & \gop & \vsinio & $i$ && \top & \gop & \vsinio & $i$ && \top & \gop & \vsinio & $i$ && \top & \gop & \vsinio & $i$\\
   &   (K)  & (cgs) & (\kps) & ($^{\rm o}$) && (K) & (cgs) & (\kps) & ($^{\rm o}$) && (K)  & (cgs) & (\kps) & ($^{\rm o}$) && (K) & (cgs) & (\kps) & ($^{\rm o}$)\\
\hline
 42406 &        &       &       &     &&        &       &       &     && 16500  &  3.96 &  338  &  61 && 17000  &  3.97 &  360  & 79\\
 43264 &        &       &       &     &&        &       &       &     &&        &       &       &     && 12000  &  3.16 &  284  &  79 \\
 43285 & 15000  &  4.03 &  266  &  71 && 15000  &  3.99 &  274  &  55 && 15000  &  3.95 &  292  &  51 && 15000  &  3.93 &  309  &  58 \\
 44783 & 12000  &  3.37 &  226  &  93 && 12000  &  3.40 &  227  &  80 && 11500  &  3.13 &  231  &  82 && 11500  &  3.21 &  218  &  57\\
 45901 & 27000  &  3.78 &  171  &  31 && 27000  &  3.79 &  173  &  26 && 27500  &  3.84 &  175  &  24 && 29500  &  4.06 &  174  &  21 \\
 46380 & 23500  &  3.91 &  306  &  65 && 23500  &  3.92 &  310  &  49 && 23500  &  3.94 &  315  &  44 && 23500  &  3.96 &  325  &  41 \\
 46484 & 27500  &  3.69 &  127  &  25 && 26500  &  3.61 &  130  &  21 && 29000  &  3.81 &  128  &  19 && 28000  &  3.69 &  134  &  15 \\
 47054 & 13000  &  3.60 &  220  &  78 && 13000  &  3.58 &  229  &  59 && 13000  &  3.51 &  239  &  54 && 12500  &  3.47 &  253  &  51 \\
 47160 & 11500  &  3.76 &  150  &  39 && 11500  &  3.74 &  158  &  33 && 11500  &  3.69 &  161  &  35 && 11500  &  3.70 &  173  &  33 \\
 47359 &        &       &       &     &&        &       &       &     && 32500  &  4.03 &  469  &  73 && 32000  &  4.01 &  486  & 71\\
 49330 & 26500  &  3.87 &  280  &  59 && 26500  &  3.86 &  285  &  46 && 28500  &  3.96 &  285  &  37 && 27000  &  3.87 &  296  &  34 \\
 49567 & 17500  &  3.58 &   93  &  26 && 17500  &  3.58 &   94  &  21 && 18000  &  3.61 &   99  &  20 && 18000  &  3.62 &   99  &  18 \\
 49585 & 26500  &  4.03 &  318  &  73 && 26000  &  4.10 &  325  &  57 && 25000  &  4.20 &  332  &  50 && 27500  &  4.09 &  331  &  41 \\
 49787 & 25500  &  4.09 &  167  &  29 && 25500  &  4.09 &  169  &  24 && 25500  &  4.10 &  172  &  22 && 25500  &  4.13 &  175  &  18 \\
 50083 & 21000  &  3.58 &  188  &  45 && 21000  &  3.61 &  193  &  37 && 21000  &  3.64 &  196  &  33 && 21500  &  3.68 &  199  &  30 \\
 50138 & 12500  &  3.27 &   67  &  21 && 12500  &  3.27 &   69  &  18 && 12500  &  3.27 &   66  &  16 && 12500  &  3.23 &   73  &  15 \\
 50209 & 13000  &  3.66 &  202  &  89 && 13000  &  3.64 &  209  &  64 && 13000  &  3.57 &  219  &  57 && 12500  &  3.53 &  230  &  55 \\
 50581 & 11000  &  3.77 &  236  &  79 && 11000  &  3.65 &  250  &  78 && 11000  &  3.66 &  251  &  64 && 10500  &  3.64 &  250  &  54 \\
 50696 &        &       &       &     && 24000  &  3.85 &  366  &  78 && 25000  &  3.88 &  366  &  61 && 23500  &  3.75 &  375  &  49 \\
 50868 & 23500  &  4.43 &  273  &  49 && 23500  &  4.47 &  276  &  39 && 23500  &  4.46 &  278  &  35 && 24000  &  4.52 &  284  &  31 \\
 50891 & 28500  &  4.00 &  227  &  45 && 28000  &  4.02 &  231  &  37 && 29000  &  4.05 &  233  &  32 && 29000  &  4.08 &  239  & 29\\
 51193 & 24000  &  3.73 &  221  &  46 && 24000  &  3.72 &  224  &  37 && 24000  &  3.74 &  228  &  34 && 24000  &  3.79 &  233  &  31 \\
 51404 & 24500  &  4.45 &  347  &  68 && 24000  &  4.45 &  353  &  52 && 24500  &  4.49 &  358  &  47 && 25000  &  4.53 &  363  &  41 \\
 51452 & 31000  &  4.02 &  306  &  59 && 30500  &  3.99 &  309  &  46 && 31500  &  4.04 &  313  &  41 && 30500  &  3.99 &  325  &  39 \\
 51506 & 18000  &  3.98 &  183  &  37 && 18000  &  4.01 &  186  &  31 && 18000  &  3.99 &  195  &  29 && 18000  &  3.98 &  204  &  28 \\
 52721 &        &       &       &     && 24000  &  4.26 &  364  &  65 && 24500  &  4.29 &  368  &  51 && 24500  &  4.33 &  377  & 45\\
 53085 & 16000  &  3.90 &  212  &  54 && 16000  &  3.88 &  222  &  45 && 16000  &  3.86 &  231  &  42 && 11500  &  3.80 &  247  &  40 \\
 53667 & 29000  &  3.56 &  116  &  23 && 27500  &  3.46 &  119  &  19 && 30000  &  3.66 &  118  &  17 && 31000  &  3.77 &  118  &  14 \\
 54464 & 17500  &  3.55 &  172  &  51 && 18000  &  3.55 &  177  &  42 && 18000  &  3.57 &  182  &  40 && 18000  &  3.56 &  191  & 38\\
 55135 & 18000  &  4.56 &  254  &  55 && 18000  &  4.55 &  264  &  55 && 18000  &  4.55 &  273  &  44 && 18000  &  4.52 &  290  & 39\\
 55606 & 26500  &  4.32 &  357  &  65 && 27000  &  4.34 &  361  &  49 && 28500  &  4.45 &  364  &  44 && 28000  &  4.43 &  377  &  41 \\
 55806 &  \\
 57539 & 14500  &  3.42 &  149  &  46 && 14000  &  3.40 &  155  &  39 && 14000  &  3.37 &  159  &  36 && 14000  &  3.32 &  171  &  34 \\
166917 & 12500  &  3.36 &  170  &  58 && 12500  &  3.35 &  173  &  46 && 12500  &  3.29 &  169  &  46 && 12500  &  3.28 &  184  &  44 \\
168797 & 20000  &  3.69 &  271  &  79 && 20500  &  3.75 &  279  &  55 && 21000  &  3.71 &  288  &  48 && 21000  &  3.74 &  302  &  45 \\
170009 & 11500  &  3.56 &  182  &  68 && 11500  &  3.45 &  181  &  53 && 11500  &  3.43 &  193  &  49 && 11000  &  3.44 &  204  &  47 \\
170714 & 24500  &  4.06 &  277  &  53 && 24500  &  4.07 &  280  &  44 && 24500  &  4.06 &  283  &  38 && 24500  &  4.10 &  290  &  35 \\
171219 &        &       &       &     && 16000  &  3.53 &  314  &  85 && 15500  &  3.53 &  314  &  83 && 16500  &  3.70 &  339  &  77 \\
173219 & 27500  &  3.73 &   63  &  11 && 27500  &  3.74 &   66  &   9 && 29000  &  3.92 &   65  &   9 && 29500  &  3.96 &   66  &   8 \\
173371 & 13500  &  3.88 &  278  &  92 && 13500  &  3.83 &  295  &  78 && 13500  &  3.77 &  313  &  69 && 13500  &  3.79 &  312  &  73 \\
173530 & 13500  &  3.43 &  239  &  94 && 13500  &  3.45 &  246  &  80 && 14000  &  3.51 &  249  &  80 && 13500  &  3.33 &  277  &  71 \\
173637 & 26500  &  3.64 &  207  &  43 && 28000  &  3.79 &  207  &  35 && 28000  &  3.78 &  208  &  29 && 29000  &  3.93 &  213  &  29 \\
173817 & 14000  &  3.94 &  267  &  76 && 14000  &  3.90 &  276  &  57 && 14000  &  3.85 &  296  &  55 && 14000  &  3.84 &  317  &  59 \\
174513 & 24000  &  3.95 &  256  &  57 && 24000  &  3.96 &  261  &  49 && 24000  &  3.98 &  264  &  43 && 24000  & 4.02  &  270  & 38\\
174571 & 22000  &  4.14 &  246  &  51 && 22000  &  4.20 &  250  &  42 && 22500  &  4.25 &  252  &  36 && 23000  & 4.31  &  259  & 30\\
174705 &        &       &       &     && 23500  &  4.10 &  341  &  69 && 23500  &  4.10 &  346  &  58 && 24000  & 4.15  &  357  & 51\\
174886 & 15000  &  3.49 &   77  &  21 && 15000  &  3.50 &   79  &  18 && 15000  &  3.49 &   79  &  17 && 15000  &  3.47 &   83  &  16 \\
175869 & 12500  &  3.52 &  168  &  58 && 12000  &  3.47 &  171  &  47 && 12000  &  3.40 &  180  &  44 && 12000  &  3.39 &  196  &  43 \\
176159 & 14500  &  3.92 &  236  &  74 && 14500  &  3.89 &  243  &  62 && 14500  &  3.89 &  258  &  58 && 14500  &  3.80 &  270  & 58\\
176630 & 17000  &  3.40 &  185  &  56 && 17000  &  3.41 &  188  &  45 && 17000  &  3.40 &  197  &  42 && 17000  &  3.37 &  211  &  40 \\
178479 & 17000  &  4.06 &  106  &  24 && 17000  &  4.08 &  109  &  20 && 17000  &  4.07 &  116  &  19 && 17000  &  4.07 &  117  & 18\\
179343 & 11500  &  3.09 &  148  &  55 && 12000  &  3.07 &  155  &  45 && 11500  &  3.02 &  158  &  43 && 11500  &  2.98 &  167  &  40 \\
179405 & 20000  &  4.19 &  237  &  44 && 20500  &  4.23 &  248  &  36 && 21000  &  4.27 &  247  &  32 && 21000  &  4.30 &  258  & 30 \\
180126 & 21000  &  3.98 &  250  &  60 && 21500  &  4.02 &  252  &  48 && 21500  &  4.09 &  259  &  42 && 21500  &  4.08 & 267  & 39\\
181231 & 14500  &  3.96 &  257  &  66 && 14500  &  3.90 &  259  &  52 && 14500  &  3.93 &  275  &  54 && 14500  &  3.83 &  295  &  54 \\
\hline
\end{tabular}
\end{table*}

\newpage

\setcounter{table}{4}

\begin{table*}
\center \caption{Continued ...}
\begin{tabular}{@{\ }ll@{\ \ }l@{\ \ }l@{\ }l@{\ \ }cl@{\ \ }l@{\ \ }l@{\ }l@{\ \ }cl@{\ \ }l@{\ \ }l@{\ }l@{\ \ }cl@{\ \ }l@{\ \ }l@{\ }l@{\ }}
\hline \hline
   & \multicolumn{4}{c}{\omc~= 0.80} && \multicolumn{4}{c}{\omc~= 0.90} && \multicolumn{4}{c}{\omc~= 0.95} && \multicolumn{4}{c}{\omc~= 0.99}\\
\cline{2-5}\cline{7-10}\cline{12-15}\cline{17-20}\\
HD &  \top & \gop & \vsinio & $i$ && \top & \gop & \vsinio & $i$ && \top & \gop & \vsinio & $i$ && \top & \gop & \vsinio & $i$\\
   &   (K)  & (cgs) & (\kps) & ($^{\rm o}$) && (K) & (cgs) & (\kps) & ($^{\rm o}$) && (K)  & (cgs) & (\kps) & ($^{\rm o}$) && (K) & (cgs) & (\kps) & ($^{\rm o}$)\\
\hline
181308 & 15000  &  4.00 &  254  &  64 && 15000  &  3.97 &  261  &  51 && 15000  &  3.93 &  277  &  48 && 15000  &  3.91 &  293  &  52 \\
181367 & 14500  &  3.94 &  282  &  78 && 14500  &  3.88 &  292  &  62 && 14500  &  3.86 &  308  &  56 && 14500  &  3.76 &  331  & 56  \\
181709 &        &       &       &     &&        &       &       &     &&        &       &       &     && 14500  &  3.40 & 291  & 90\\
181803 &        &       &       &     && 13000  &  3.19 &  190  &  79 && 12500  &  3.08 &  186  &  66 && 12500  &  3.11 & 210  & 62\\
184279 & 30500  &  4.05 &  135  &  22 && 31000  &  4.08 &  137  &  19 && 31000  &  4.10 &  140  &  17 && 31000  &  4.08 &  145  &  16 \\
184767 & 10000  &  3.51 &   49  &  16 && 10500  &  3.55 &   49  &  13 && 10000  &  3.43 &   46  &  13 && 10000  & 3.53  & 50   & 10\\
194244 & 11500  &  3.59 &  234  &  84 && 11000  &  3.52 &  232  &  79 && 11000  &  3.40 &  256  &  70 && 11000  &  3.42 &  247  &  61 \\
230579 & 31000  &  4.03 &  338  & 72  && 30000  &  3.92 &  343  &  47 && 32000  &  4.07 & 346   &  47 && 31500  &  4.08 & 352 & 38\\
BD-09\d4858 & 21500 & 4.18 & 115 & 22 && 22000  &  4.22 &  115  &  17 && 22000  &  4.27 &  117  &  15 && 22000  &  4.31 & 117  & 14\\
\hline
\end{tabular}
\end{table*}

\newpage


\setlongtables \onecolumn

\setcounter{table}{6} \setlength{\LTcapwidth}{17cm}

\begin{longtable}{rrrrrrr}
\kill


\caption{List of $pnrc$ surface-$averaged$ parameters and their
respective interpolated masses $M/M_{\odot}$, ages $\tau$, and
fractional ages $\tau/\tau_{\rm MS}$. When the projected rotation
velocity was greater than the break-up speed or than the
equatorial speed, the Table were left blank. Can be downloaded
from the CDS.}

\\
\hline \hline\noalign{\smallskip}
 HD & \teff $^{\rm surf.}$ & \logg $^{\rm surf.}$ & $\log
 L/L_\odot$ $^{\rm surf.}$ & M/M$_\odot$ & age (years) & $\tau / \tau_{\rm
 MS}$\\ \noalign{\smallskip}
 \hline
\multicolumn{7}{|c|}{\omc\ = 0.80}\\
 \hline
 42406 \\
 43264 \\
 43285 & 14500$\pm$1100 & 4.02$\pm$0.17 &  2.69$\pm$ 0.09 &  4.60$\pm$ 0.30 & 0.665E+08$\pm$0.201E+08 &  0.47$\pm$ 0.13 \\
 44783 & 11500$\pm$1100 & 3.33$\pm$0.21 &  2.94$\pm$ 0.12 &  4.30$\pm$ 0.30 & 0.166E+09$\pm$0.155E+08 &  1.02$\pm$ 0.03 \\
 45901 & 26000$\pm$2300 & 3.74$\pm$0.21 &  4.43$\pm$ 0.13 & 13.60$\pm$ 1.30 & 0.122E+08$\pm$0.200E+07 &  0.75$\pm$ 0.10 \\
 46380 & 23000$\pm$1100 & 3.88$\pm$0.14 &  3.94$\pm$ 0.12 & 10.00$\pm$ 0.60 & 0.166E+08$\pm$0.256E+07 &  0.62$\pm$ 0.09 \\
 46484 & 27000$\pm$1000 & 3.64$\pm$0.13 &  4.65$\pm$ 0.14 & 15.70$\pm$ 1.30 & 0.110E+08$\pm$0.604E+06 &  0.80$\pm$ 0.04 \\
 47054 & 12500$\pm$ 800 & 3.58$\pm$0.15 &  2.87$\pm$ 0.12 &  4.60$\pm$ 0.20 & 0.132E+09$\pm$0.110E+08 &  0.91$\pm$ 0.05 \\
 47160 & 11500$\pm$ 600 & 3.74$\pm$0.11 &  2.41$\pm$ 0.08 &  3.60$\pm$ 0.10 & 0.208E+09$\pm$0.169E+08 &  0.79$\pm$ 0.05 \\
47359\\
 49330 & 26000$\pm$1700 & 3.84$\pm$0.17 &  4.30$\pm$ 0.12 & 12.80$\pm$ 1.00 & 0.118E+08$\pm$0.195E+07 &  0.69$\pm$ 0.10 \\
 49567 & 17000$\pm$1700 & 3.56$\pm$0.22 &  3.58$\pm$ 0.12 &  7.00$\pm$ 0.60 & 0.477E+08$\pm$0.646E+07 &  0.91$\pm$ 0.07 \\
 49585 & 26000$\pm$1700 & 4.00$\pm$0.16 &  4.11$\pm$ 0.09 & 12.00$\pm$ 0.90 & 0.876E+07$\pm$0.265E+07 &  0.45$\pm$ 0.12 \\
 49787 & 25000$\pm$1200 & 4.07$\pm$0.12 &  3.93$\pm$ 0.08 & 10.90$\pm$ 0.60 & 0.796E+07$\pm$0.250E+07 &  0.34$\pm$ 0.10 \\
 50083 & 20500$\pm$1100 & 3.56$\pm$0.15 &  4.04$\pm$ 0.12 &  9.60$\pm$ 0.60 & 0.255E+08$\pm$0.191E+07 &  0.89$\pm$ 0.04 \\
 50138 & 12500$\pm$1100 & 3.23$\pm$0.19 &  3.22$\pm$ 0.13 &  5.10$\pm$ 0.40 & 0.113E+09$\pm$0.100E+08 &  1.02$\pm$ 0.02 \\
 50209 & 12750$\pm$1700 & 3.63$\pm$0.28 &  2.82$\pm$ 0.14 &  4.50$\pm$ 0.40 & 0.133E+09$\pm$0.262E+08 &  0.87$\pm$ 0.11 \\
 50581 & 10500$\pm$ 600 & 3.74$\pm$0.11 &  2.26$\pm$ 0.07 &  3.30$\pm$ 0.10 & 0.260E+09$\pm$0.227E+08 &  0.78$\pm$ 0.05 \\
50696\\
 50868 & 22500$\pm$1600 & 4.42$\pm$0.15 &  3.29$\pm$ 0.07 &  8.10$\pm$ 0.20 &   &    \\
 50891 & 27500$\pm$2300 & 3.97$\pm$0.22 &  4.29$\pm$ 0.16 & 13.60$\pm$ 1.50 & 0.816E+07$\pm$0.239E+07 &  0.49$\pm$ 0.14 \\
 51193 & 23000$\pm$1200 & 3.70$\pm$0.12 &  4.19$\pm$ 0.09 & 11.10$\pm$ 0.50 & 0.173E+08$\pm$0.131E+07 &  0.79$\pm$ 0.05 \\
 51404 & 23500$\pm$1900 & 4.44$\pm$0.14 &  3.37$\pm$ 0.10 &  8.70$\pm$ 0.80 &   &    \\
 51452 & 30000$\pm$1700 & 3.99$\pm$0.16 &  4.51$\pm$ 0.12 & 16.40$\pm$ 1.40 & 0.571E+07$\pm$0.162E+07 &  0.43$\pm$ 0.12 \\
 51506 & 17500$\pm$1100 & 3.96$\pm$0.17 &  3.17$\pm$ 0.12 &  6.10$\pm$ 0.40 & 0.393E+08$\pm$0.948E+07 &  0.55$\pm$ 0.12 \\
52721\\
 53085 & 15500$\pm$ 800 & 3.87$\pm$0.12 &  2.99$\pm$ 0.08 &  5.30$\pm$ 0.20 & 0.658E+08$\pm$0.891E+07 &  0.66$\pm$ 0.07 \\
 53667 & 28000$\pm$1700 & 3.51$\pm$0.14 &  4.96$\pm$ 0.09 & 19.30$\pm$ 1.40 & 0.898E+07$\pm$0.566E+06 &  0.87$\pm$ 0.03 \\
 54464 & 17000$\pm$1700 & 3.52$\pm$0.22 &  3.67$\pm$ 0.13 &  7.30$\pm$ 0.60 & 0.440E+08$\pm$0.545E+07 &  0.92$\pm$ 0.06 \\
   55135 & 17618$\pm$1100 & 4.27$\pm$0.21 &  2.84$\pm$0.20 &  5.64$\pm$0.46 &  0.134E+08$\pm$0.660E+07 & 0.16$\pm$0.08 \\
 55606 & 26000$\pm$2300 & 4.30$\pm$0.21 &  3.76$\pm$ 0.13 & 10.80$\pm$ 1.30 & 0.180E+06$\pm$0.293E+07 &  0.01$\pm$ 0.14 \\
 55806 &  \\
 57539 & 14000$\pm$1100 & 3.39$\pm$0.18 &  3.31$\pm$ 0.10 &  5.50$\pm$ 0.40 & 0.907E+08$\pm$0.802E+07 &  1.00$\pm$ 0.03 \\
166917 & 12000$\pm$1100 & 3.32$\pm$0.20 &  3.08$\pm$ 0.12 &  4.70$\pm$ 0.30 & 0.136E+09$\pm$0.119E+08 &  1.02$\pm$ 0.03 \\
168797 & 19500$\pm$1100 & 3.66$\pm$0.15 &  3.78$\pm$ 0.12 &  8.20$\pm$ 0.50 & 0.319E+08$\pm$0.283E+07 &  0.82$\pm$ 0.06 \\
170009 & 11500$\pm$ 600 & 3.54$\pm$0.17 &  2.65$\pm$ 0.16 &  4.00$\pm$ 0.20 & 0.192E+09$\pm$0.156E+08 &  0.93$\pm$ 0.05 \\
170714 & 24000$\pm$1400 & 4.04$\pm$0.16 &  3.86$\pm$ 0.12 & 10.20$\pm$ 0.70 & 0.104E+08$\pm$0.327E+07 &  0.39$\pm$ 0.12 \\
171219\\
173219 & 26500$\pm$2300 & 3.69$\pm$0.22 &  4.56$\pm$ 0.18 & 14.80$\pm$ 1.70 & 0.116E+08$\pm$0.168E+07 &  0.79$\pm$ 0.10 \\
173371 & 13000$\pm$ 600 & 3.86$\pm$0.11 &  2.58$\pm$ 0.08 &  4.10$\pm$ 0.20 & 0.127E+09$\pm$0.130E+08 &  0.67$\pm$ 0.06 \\
173530 & 13000$\pm$1100 & 3.40$\pm$0.19 &  3.16$\pm$ 0.10 &  5.10$\pm$ 0.30 & 0.111E+09$\pm$0.106E+08 &  1.00$\pm$ 0.03 \\
173637 & 25500$\pm$1700 & 3.60$\pm$0.15 &  4.57$\pm$ 0.10 & 14.40$\pm$ 1.00 & 0.128E+08$\pm$0.101E+07 &  0.85$\pm$ 0.05 \\
173817 & 14000$\pm$1100 & 3.92$\pm$0.19 &  2.67$\pm$ 0.12 &  4.40$\pm$ 0.30 & 0.953E+08$\pm$0.229E+08 &  0.61$\pm$ 0.12 \\
174513 & 23500$\pm$1700 & 3.92$\pm$0.17 &  3.95$\pm$ 0.10 & 10.30$\pm$ 0.80 & 0.146E+08$\pm$0.356E+07 &  0.57$\pm$ 0.12 \\
174571 & 21000$\pm$1700 & 4.12$\pm$0.19 &  3.47$\pm$ 0.10 &  8.10$\pm$ 0.60 & 0.108E+08$\pm$0.602E+07 &  0.26$\pm$ 0.13 \\
174705\\
174886 & 14500$\pm$ 600 & 3.47$\pm$0.10 &  3.32$\pm$ 0.08 &  5.80$\pm$ 0.20 & 0.796E+08$\pm$0.432E+07 &  0.97$\pm$ 0.02 \\
175869 & 12000$\pm$ 600 & 3.49$\pm$0.11 &  2.86$\pm$ 0.08 &  4.40$\pm$ 0.20 & 0.151E+09$\pm$0.992E+07 &  0.96$\pm$ 0.03 \\
176159 & 14500$\pm$1100 & 3.90$\pm$0.18 &  2.76$\pm$ 0.10 &  4.60$\pm$ 0.30 & 0.868E+08$\pm$0.184E+08 &  0.63$\pm$ 0.11 \\
176630 & 16500$\pm$1100 & 3.36$\pm$0.17 &  3.74$\pm$ 0.13 &  7.20$\pm$ 0.60 & 0.491E+08$\pm$0.429E+07 &  1.00$\pm$ 0.03 \\
178479 & 16500$\pm$1100 & 4.04$\pm$0.16 &  2.94$\pm$ 0.08 &  5.50$\pm$ 0.30 & 0.399E+08$\pm$0.131E+08 &  0.43$\pm$ 0.12 \\
179343 & 11500$\pm$1100 & 3.05$\pm$0.22 &  3.27$\pm$ 0.18 &  5.20$\pm$ 0.50 & 0.106E+09$\pm$0.126E+08 &  1.02$\pm$ 0.00 \\
  179405 & 19500$\pm$2300 & 4.17$\pm$0.27 &  3.19$\pm$0.16 &  6.87$\pm$0.77 &  0.186E+08$\pm$0.705E+07 & 0.28$\pm$0.10 \\
180126 & 20000$\pm$1700 & 3.96$\pm$0.19 &  3.53$\pm$ 0.10 &  7.80$\pm$ 0.60 & 0.230E+08$\pm$0.672E+07 &  0.53$\pm$ 0.13 \\
181231 & 14000$\pm$1100 & 3.94$\pm$0.18 &  2.71$\pm$ 0.10 &  4.60$\pm$ 0.30 & 0.844E+08$\pm$0.210E+08 &  0.58$\pm$ 0.12 \\
181308 & 14500$\pm$1100 & 3.98$\pm$0.17 &  2.74$\pm$ 0.10 &  4.70$\pm$ 0.30 & 0.697E+08$\pm$0.197E+08 &  0.52$\pm$ 0.13 \\
  181367 & 14500$\pm$1100 & 3.92$\pm$0.18 &  2.75$\pm$0.12 &  4.67$\pm$0.30 &  0.802E+08$\pm$0.136E+08 & 0.57$\pm$0.08 \\
181709\\
181803\\
184279 & 30000$\pm$2300 & 4.02$\pm$0.22 &  4.45$\pm$ 0.17 & 15.90$\pm$ 1.80 & 0.540E+07$\pm$0.203E+07 &  0.39$\pm$ 0.14 \\
184767 & 10000$\pm$ 600 & 3.49$\pm$0.13 &  2.39$\pm$ 0.08 &  3.40$\pm$ 0.20 & 0.309E+09$\pm$0.285E+08 &  0.97$\pm$ 0.03 \\
194244 & 11000$\pm$ 600 & 3.57$\pm$0.11 &  2.54$\pm$ 0.07 &  3.80$\pm$ 0.10 & 0.217E+09$\pm$0.151E+08 &  0.91$\pm$ 0.04 \\
230579 & 30000$\pm$1700 & 4.00$\pm$0.16 &  4.49$\pm$0.12 & 16.33$\pm$1.36 &  0.544E+07$\pm$0.116E+07 & 0.40$\pm$0.08 \\
BD-09\d4858 & 21000$\pm$1700 & 4.16$\pm$0.21 &  3.37$\pm$ 0.15 &  7.60$\pm$ 0.70 & 0.884E+07$\pm$0.636E+07 &  0.19$\pm$ 0.13 \\
\hline
\multicolumn{7}{|c|}{\omc\ = 0.90}\\
 \hline
 42406\\
 43264\\
 43285 & 14500$\pm$1100 & 3.95$\pm$0.18 &  2.70$\pm$ 0.10 &  4.60$\pm$ 0.30 & 0.819E+08$\pm$0.201E+08 &  0.57$\pm$ 0.12 \\
 44783 & 11500$\pm$1100 & 3.35$\pm$0.20 &  2.94$\pm$ 0.12 &  4.40$\pm$ 0.30 & 0.161E+09$\pm$0.151E+08 &  1.02$\pm$ 0.03 \\
 45901 & 26000$\pm$2200 & 3.74$\pm$0.21 &  4.42$\pm$ 0.13 & 13.50$\pm$ 1.30 & 0.125E+08$\pm$0.196E+07 &  0.75$\pm$ 0.10 \\
 46380 & 22500$\pm$1100 & 3.87$\pm$0.14 &  3.91$\pm$ 0.12 &  9.80$\pm$ 0.60 & 0.177E+08$\pm$0.264E+07 &  0.64$\pm$ 0.09 \\
 46484 & 25500$\pm$1000 & 3.54$\pm$0.13 &  4.65$\pm$ 0.14 & 15.10$\pm$ 1.30 & 0.124E+08$\pm$0.606E+06 &  0.87$\pm$ 0.03 \\
 47054 & 12500$\pm$ 800 & 3.54$\pm$0.16 &  2.89$\pm$ 0.12 &  4.60$\pm$ 0.20 & 0.133E+09$\pm$0.117E+08 &  0.93$\pm$ 0.05 \\
 47160 & 11000$\pm$ 600 & 3.69$\pm$0.11 &  2.44$\pm$ 0.07 &  3.70$\pm$ 0.10 & 0.211E+09$\pm$0.154E+08 &  0.83$\pm$ 0.05 \\
 47359\\
 49330 & 25500$\pm$1700 & 3.80$\pm$0.17 &  4.30$\pm$ 0.12 & 12.60$\pm$ 1.00 & 0.127E+08$\pm$0.190E+07 &  0.74$\pm$ 0.09 \\
 49567 & 17000$\pm$1700 & 3.54$\pm$0.22 &  3.57$\pm$ 0.12 &  7.00$\pm$ 0.50 & 0.488E+08$\pm$0.638E+07 &  0.92$\pm$ 0.06 \\
 49585 & 25000$\pm$1700 & 4.05$\pm$0.16 &  3.95$\pm$ 0.09 & 11.00$\pm$ 0.80 & 0.847E+07$\pm$0.317E+07 &  0.37$\pm$ 0.12 \\
 49787 & 24500$\pm$1100 & 4.04$\pm$0.12 &  3.92$\pm$ 0.08 & 10.70$\pm$ 0.60 & 0.936E+07$\pm$0.257E+07 &  0.39$\pm$ 0.10 \\
 50083 & 20500$\pm$1100 & 3.55$\pm$0.15 &  4.02$\pm$ 0.12 &  9.40$\pm$ 0.60 & 0.253E+08$\pm$0.189E+07 &  0.86$\pm$ 0.04 \\
 50138 & 12000$\pm$1200 & 3.21$\pm$0.19 &  3.21$\pm$ 0.13 &  5.10$\pm$ 0.40 & 0.113E+09$\pm$0.988E+07 &  1.02$\pm$ 0.02 \\
 50209 & 12500$\pm$1700 & 3.60$\pm$0.28 &  2.83$\pm$ 0.14 &  4.50$\pm$ 0.40 & 0.135E+09$\pm$0.250E+08 &  0.90$\pm$ 0.10 \\
 50581 & 10500$\pm$ 600 & 3.61$\pm$0.12 &  2.37$\pm$ 0.07 &  3.50$\pm$ 0.10 & 0.263E+09$\pm$0.200E+08 &  0.89$\pm$ 0.04 \\
 50696 & 23500$\pm$1700 & 3.80$\pm$0.18 &  4.06$\pm$ 0.12 & 10.60$\pm$ 0.80 & 0.169E+08$\pm$0.273E+07 &  0.70$\pm$ 0.09 \\
 50868 & 22500$\pm$1100 & 4.43$\pm$0.39 &  3.24$\pm$ 0.43 &  7.90$\pm$ 1.10 &   &    \\
 50891 & 27000$\pm$2200 & 3.97$\pm$0.22 &  4.25$\pm$ 0.16 & 13.20$\pm$ 1.40 & 0.839E+07$\pm$0.263E+07 &  0.49$\pm$ 0.14 \\
 51193 & 23000$\pm$1100 & 3.66$\pm$0.12 &  4.19$\pm$ 0.09 & 11.10$\pm$ 0.50 & 0.180E+08$\pm$0.126E+07 &  0.81$\pm$ 0.04 \\
 51404 & 23500$\pm$1900 & 4.42$\pm$0.15 &  3.35$\pm$ 0.09 &  8.50$\pm$ 0.70 &   &    \\
 51452 & 29500$\pm$1700 & 3.94$\pm$0.16 &  4.50$\pm$ 0.12 & 15.80$\pm$ 1.30 & 0.714E+07$\pm$0.170E+07 &  0.52$\pm$ 0.14 \\
 51506 & 17000$\pm$1100 & 3.97$\pm$0.17 &  3.13$\pm$ 0.12 &  6.00$\pm$ 0.40 & 0.402E+08$\pm$0.101E+08 &  0.54$\pm$ 0.12 \\
 52721 & 23500$\pm$2200 & 4.23$\pm$0.27 &  3.57$\pm$ 0.22 &  9.10$\pm$ 1.00 & 0.262E+07$\pm$0.465E+07 &  0.08$\pm$ 0.14 \\
 53085 & 15500$\pm$ 800 & 3.83$\pm$0.12 &  3.00$\pm$ 0.08 &  5.30$\pm$ 0.20 & 0.707E+08$\pm$0.839E+07 &  0.70$\pm$ 0.07 \\
 53667 & 26500$\pm$1700 & 3.38$\pm$0.14 &  4.97$\pm$ 0.09 & 18.90$\pm$ 1.40 & 0.972E+07$\pm$0.561E+06 &  0.92$\pm$ 0.03 \\
 54464 & 17000$\pm$1700 & 3.50$\pm$0.22 &  3.67$\pm$ 0.13 &  7.40$\pm$ 0.60 & 0.443E+08$\pm$0.606E+07 &  0.93$\pm$ 0.06 \\
   55135& 17500$\pm$1100 &  4.25$\pm$0.21 &  2.83$\pm$0.20 &  5.59$\pm$0.45&   0.156E+08$\pm$0.707E+07& 0.18$\pm$0.08\\
 55606 & 26000$\pm$2200 & 4.31$\pm$0.21 &  3.77$\pm$ 0.14 & 10.90$\pm$ 1.30 & 0.136E+06$\pm$0.274E+07 &  0.01$\pm$ 0.14 \\
 55806 & 13000$\pm$1100 & 3.58$\pm$0.19 &  2.90$\pm$ 0.10 &  4.70$\pm$ 0.30 & 0.125E+09$\pm$0.150E+08 &  0.91$\pm$ 0.06 \\
 57539 & 13500$\pm$1100 & 3.34$\pm$0.18 &  3.31$\pm$ 0.10 &  5.50$\pm$ 0.40 & 0.938E+08$\pm$0.767E+07 &  1.01$\pm$ 0.03 \\
166917 & 12000$\pm$1100 & 3.29$\pm$0.20 &  3.07$\pm$ 0.13 &  4.70$\pm$ 0.30 & 0.137E+09$\pm$0.117E+08 &  1.02$\pm$ 0.03 \\
168797 & 19500$\pm$1100 & 3.70$\pm$0.15 &  3.76$\pm$ 0.12 &  8.30$\pm$ 0.50 & 0.304E+08$\pm$0.289E+07 &  0.80$\pm$ 0.06 \\
170009 & 11000$\pm$ 600 & 3.40$\pm$0.17 &  2.71$\pm$ 0.16 &  3.90$\pm$ 0.20 & 0.215E+09$\pm$0.151E+08 &  1.00$\pm$ 0.03 \\
170714 & 23500$\pm$1400 & 4.03$\pm$0.16 &  3.82$\pm$ 0.12 &  9.90$\pm$ 0.70 & 0.115E+08$\pm$0.347E+07 &  0.42$\pm$ 0.12 \\
171219 & 15500$\pm$ 800 & 3.48$\pm$0.14 &  3.42$\pm$ 0.12 &  6.20$\pm$ 0.30 & 0.658E+08$\pm$0.471E+07 &  0.96$\pm$ 0.03 \\
173219 & 26500$\pm$2200 & 3.68$\pm$0.22 &  4.55$\pm$ 0.18 & 14.60$\pm$ 1.70 & 0.119E+08$\pm$0.170E+07 &  0.79$\pm$ 0.10 \\
173371 & 13000$\pm$ 600 & 3.79$\pm$0.10 &  2.70$\pm$ 0.08 &  4.40$\pm$ 0.20 & 0.122E+09$\pm$0.978E+07 &  0.75$\pm$ 0.05 \\
173530 & 13000$\pm$1100 & 3.40$\pm$0.18 &  3.15$\pm$ 0.10 &  5.10$\pm$ 0.30 & 0.111E+09$\pm$0.104E+08 &  1.00$\pm$ 0.03 \\
173637 & 27000$\pm$1700 & 3.73$\pm$0.15 &  4.53$\pm$ 0.09 & 14.70$\pm$ 1.00 & 0.111E+08$\pm$0.119E+07 &  0.75$\pm$ 0.06 \\
173817 & 13500$\pm$1100 & 3.85$\pm$0.19 &  2.68$\pm$ 0.12 &  4.40$\pm$ 0.30 & 0.109E+09$\pm$0.213E+08 &  0.68$\pm$ 0.11 \\
174513 & 23000$\pm$1700 & 3.92$\pm$0.17 &  3.91$\pm$ 0.10 & 10.00$\pm$ 0.80 & 0.156E+08$\pm$0.368E+07 &  0.58$\pm$ 0.12 \\
174571 & 21000$\pm$1700 & 4.16$\pm$0.19 &  3.42$\pm$ 0.10 &  8.00$\pm$ 0.60 & 0.835E+07$\pm$0.584E+07 &  0.20$\pm$ 0.12 \\
174705 & 22500$\pm$1700 & 4.06$\pm$0.18 &  3.68$\pm$ 0.10 &  9.10$\pm$ 0.70 & 0.122E+08$\pm$0.479E+07 &  0.38$\pm$ 0.13 \\
174886 & 14500$\pm$ 600 & 3.46$\pm$0.10 &  3.30$\pm$ 0.08 &  5.70$\pm$ 0.20 & 0.816E+08$\pm$0.441E+07 &  0.97$\pm$ 0.02 \\
175869 & 12000$\pm$ 600 & 3.43$\pm$0.12 &  2.86$\pm$ 0.08 &  4.30$\pm$ 0.20 & 0.164E+09$\pm$0.117E+08 &  0.99$\pm$ 0.02 \\
176159 & 14000$\pm$1100 & 3.85$\pm$0.18 &  2.76$\pm$ 0.10 &  4.60$\pm$ 0.30 & 0.973E+08$\pm$0.181E+08 &  0.68$\pm$ 0.10 \\
176630 & 16000$\pm$1100 & 3.35$\pm$0.17 &  3.71$\pm$ 0.13 &  7.10$\pm$ 0.50 & 0.509E+08$\pm$0.430E+07 &  1.00$\pm$ 0.03 \\
178479 & 16500$\pm$1100 & 4.04$\pm$0.16 &  2.91$\pm$ 0.09 &  5.40$\pm$ 0.30 & 0.418E+08$\pm$0.132E+08 &  0.43$\pm$ 0.12 \\
179343 & 11500$\pm$1100 & 3.01$\pm$0.21 &  3.30$\pm$ 0.18 &  5.30$\pm$ 0.60 & 0.996E+08$\pm$0.126E+08 &  1.02$\pm$ 0.00 \\
  179405& 19500$\pm$2200 &  4.19$\pm$0.27 &  3.18$\pm$0.16 &  6.92$\pm$0.77&   0.172E+08$\pm$0.665E+07& 0.27$\pm$0.10\\
180126 & 20500$\pm$1700 & 3.97$\pm$0.19 &  3.54$\pm$ 0.10 &  7.90$\pm$ 0.60 & 0.214E+08$\pm$0.646E+07 &  0.51$\pm$ 0.13 \\
181231 & 13500$\pm$1100 & 3.86$\pm$0.18 &  2.72$\pm$ 0.10 &  4.50$\pm$ 0.30 & 0.102E+09$\pm$0.195E+08 &  0.68$\pm$ 0.10 \\
181308 & 14500$\pm$1100 & 3.93$\pm$0.17 &  2.74$\pm$ 0.10 &  4.70$\pm$ 0.30 & 0.818E+08$\pm$0.190E+08 &  0.59$\pm$ 0.12 \\
  181367& 14000$\pm$1100 &  3.84$\pm$0.18 &  2.77$\pm$0.12 &  4.64$\pm$0.29&   0.943E+08$\pm$0.1261E+08& 0.66$\pm$0.07\\
181709\\
181803 & 12500$\pm$ 600 & 3.13$\pm$0.19 &  3.36$\pm$ 0.22 &  5.50$\pm$ 0.70 & 0.914E+08$\pm$0.134E+08 &  1.02$\pm$ 0.01 \\
184279 & 30000$\pm$2300 & 4.03$\pm$0.22 &  4.44$\pm$ 0.18 & 15.80$\pm$ 1.80 & 0.534E+07$\pm$0.204E+07 &  0.38$\pm$ 0.14 \\
184767 & 10000$\pm$ 600 & 3.51$\pm$0.13 &  2.37$\pm$ 0.08 &  3.40$\pm$ 0.10 & 0.304E+09$\pm$0.220E+08 &  0.96$\pm$ 0.03 \\
194244 & 10500$\pm$ 600 & 3.48$\pm$0.12 &  2.58$\pm$ 0.07 &  3.80$\pm$ 0.20 & 0.231E+09$\pm$0.192E+08 &  0.97$\pm$ 0.03 \\
  230579& 28500$\pm$1700 &  3.87$\pm$0.16 &  4.53$\pm$0.12 & 15.76$\pm$1.29&   0.804E+07$\pm$0.106E+07& 0.58$\pm$0.07\\
BD-09\d4858 & 21000$\pm$1700 & 4.18$\pm$0.23 &  3.36$\pm$ 0.19 &  7.70$\pm$ 0.70 & 0.651E+07$\pm$0.595E+07 &  0.14$\pm$ 0.13 \\
 \hline
\multicolumn{7}{|c|}{\omc\ = 0.99}\\
 \hline
 42406 & 16000$\pm$1100 & 3.91$\pm$0.16 &  3.01$\pm$ 0.09 &  5.50$\pm$ 0.30 & 0.574E+08$\pm$0.114E+08 &  0.62$\pm$ 0.10 \\
 43264 & 11500$\pm$ 600 & 3.08$\pm$0.11 &  3.21$\pm$ 0.09 &  5.10$\pm$ 0.30 & 0.112E+09$\pm$0.739E+07 &  1.02$\pm$ 0.00 \\
 43285 & 14000$\pm$1100 & 3.87$\pm$0.17 &  2.77$\pm$ 0.10 &  4.70$\pm$ 0.30 & 0.919E+08$\pm$0.173E+08 &  0.67$\pm$ 0.10 \\
 44783 & 11000$\pm$1100 & 3.13$\pm$0.21 &  3.04$\pm$ 0.14 &  4.60$\pm$ 0.30 & 0.143E+09$\pm$0.128E+08 &  1.02$\pm$ 0.01 \\
 45901 & 28000$\pm$2200 & 3.98$\pm$0.20 &  4.34$\pm$ 0.12 & 14.30$\pm$ 1.40 & 0.733E+07$\pm$0.233E+07 &  0.47$\pm$ 0.14 \\
 46380 & 22500$\pm$1100 & 3.89$\pm$0.14 &  3.87$\pm$ 0.12 &  9.60$\pm$ 0.60 & 0.177E+08$\pm$0.273E+07 &  0.62$\pm$ 0.09 \\
 46484 & 26500$\pm$1000 & 3.60$\pm$0.13 &  4.66$\pm$ 0.14 & 15.70$\pm$ 1.30 & 0.113E+08$\pm$0.580E+06 &  0.83$\pm$ 0.04 \\
 47054 & 12000$\pm$ 800 & 3.40$\pm$0.16 &  2.92$\pm$ 0.12 &  4.40$\pm$ 0.30 & 0.155E+09$\pm$0.133E+08 &  1.00$\pm$ 0.03 \\
 47160 & 11000$\pm$ 600 & 3.63$\pm$0.12 &  2.45$\pm$ 0.08 &  3.60$\pm$ 0.10 & 0.226E+09$\pm$0.177E+08 &  0.88$\pm$ 0.04 \\
 47359 & 30500$\pm$2200 & 3.93$\pm$0.19 &  4.63$\pm$ 0.13 & 17.70$\pm$ 1.80 & 0.625E+07$\pm$0.166E+07 &  0.52$\pm$ 0.13 \\
 49330 & 26000$\pm$1700 & 3.79$\pm$0.17 &  4.33$\pm$ 0.12 & 13.00$\pm$ 1.00 & 0.123E+08$\pm$0.176E+07 &  0.71$\pm$ 0.09 \\
 49567 & 17000$\pm$1700 & 3.54$\pm$0.22 &  3.60$\pm$ 0.12 &  7.10$\pm$ 0.60 & 0.463E+08$\pm$0.595E+07 &  0.92$\pm$ 0.06 \\
 49585 & 26000$\pm$1700 & 4.02$\pm$0.16 &  4.10$\pm$ 0.09 & 12.10$\pm$ 0.90 & 0.813E+07$\pm$0.256E+07 &  0.43$\pm$ 0.13 \\
 49787 & 24500$\pm$1100 & 4.06$\pm$0.12 &  3.87$\pm$ 0.08 & 10.40$\pm$ 0.60 & 0.928E+07$\pm$0.274E+07 &  0.37$\pm$ 0.10 \\
 50083 & 20500$\pm$1100 & 3.59$\pm$0.14 &  4.00$\pm$ 0.12 &  9.40$\pm$ 0.60 & 0.247E+08$\pm$0.190E+07 &  0.84$\pm$ 0.04 \\
 50138 & 12000$\pm$1100 & 3.15$\pm$0.19 &  3.26$\pm$ 0.13 &  5.20$\pm$ 0.40 & 0.104E+09$\pm$0.894E+07 &  1.02$\pm$ 0.01 \\
 50209 & 12000$\pm$1700 & 3.46$\pm$0.29 &  2.88$\pm$ 0.15 &  4.40$\pm$ 0.40 & 0.152E+09$\pm$0.233E+08 &  0.98$\pm$ 0.07 \\
 50581 & 10000$\pm$ 600 & 3.58$\pm$0.12 &  2.26$\pm$ 0.07 &  3.20$\pm$ 0.10 & 0.320E+09$\pm$0.288E+08 &  0.91$\pm$ 0.04 \\
 50696 & 22500$\pm$1700 & 3.67$\pm$0.18 &  4.12$\pm$ 0.12 & 10.60$\pm$ 0.80 & 0.196E+08$\pm$0.229E+07 &  0.80$\pm$ 0.07 \\
 50868 & 22500$\pm$1100 & 4.46$\pm$0.33 &  3.22$\pm$ 0.36 &  8.00$\pm$ 1.00 & & \\
 50891 & 27500$\pm$2200 & 4.00$\pm$0.23 &  4.26$\pm$ 0.18 & 13.60$\pm$ 1.50 & 0.731E+07$\pm$0.256E+07 &  0.44$\pm$ 0.15 \\
 51193 & 23000$\pm$1100 & 3.71$\pm$0.12 &  4.14$\pm$ 0.09 & 10.90$\pm$ 0.50 & 0.178E+08$\pm$0.146E+07 &  0.77$\pm$ 0.05 \\
 51404 & 23500$\pm$1700 & 4.47$\pm$0.16 &  3.31$\pm$ 0.07 &  8.50$\pm$ 0.60 & & \\
 51452 & 29000$\pm$1700 & 3.91$\pm$0.16 &  4.51$\pm$ 0.13 & 15.80$\pm$ 1.30 & 0.762E+07$\pm$0.161E+07 &  0.56$\pm$ 0.11 \\
 51506 & 17000$\pm$1100 & 3.91$\pm$0.16 &  3.18$\pm$ 0.12 &  6.10$\pm$ 0.40 & 0.435E+08$\pm$0.870E+07 &  0.61$\pm$ 0.11 \\
 52721 & 23000$\pm$2200 & 4.27$\pm$0.23 &  3.50$\pm$ 0.14 &  8.80$\pm$ 1.00 & 0.123E+07$\pm$0.436E+07 &  0.03$\pm$ 0.14 \\
 53085 & 15000$\pm$ 800 & 3.73$\pm$0.12 &  3.05$\pm$ 0.08 &  5.30$\pm$ 0.20 & 0.793E+08$\pm$0.698E+07 &  0.80$\pm$ 0.05 \\
 53667 & 29500$\pm$1700 & 3.69$\pm$0.13 &  4.85$\pm$ 0.09 & 18.90$\pm$ 1.30 & 0.819E+07$\pm$0.712E+06 &  0.76$\pm$ 0.05 \\
 54464 & 17000$\pm$1700 & 3.49$\pm$0.22 &  3.68$\pm$ 0.13 &  7.40$\pm$ 0.60 & 0.442E+08$\pm$0.585E+07 &  0.94$\pm$ 0.06 \\
   55135 & 17000$\pm$1100 & 4.22$\pm$0.22 & 2.82$\pm$0.21 & 5.53$\pm$0.46 &  0.185E+08$\pm$0.767E+07 & 0.21$\pm$0.09\\
 55606 & 26500$\pm$1300 & 4.37$\pm$0.23 &  3.74$\pm$0.14 &  11.0$\pm$0.50 & &  \\
 55806 & 12500$\pm$1100 & 3.43$\pm$0.19 &  2.95$\pm$ 0.10 &  4.60$\pm$ 0.30 & 0.141E+09$\pm$0.143E+08 &  0.99$\pm$ 0.04 \\
 57539 & 13500$\pm$1100 & 3.24$\pm$0.17 &  3.38$\pm$ 0.10 &  5.70$\pm$ 0.30 & 0.865E+08$\pm$0.635E+07 &  1.02$\pm$ 0.01 \\
166917 & 11500$\pm$1100 & 3.20$\pm$0.20 &  3.12$\pm$ 0.13 &  4.80$\pm$ 0.30 & 0.127E+09$\pm$0.110E+08 &  1.02$\pm$ 0.01 \\
168797 & 20000$\pm$1100 & 3.67$\pm$0.15 &  3.81$\pm$ 0.12 &  8.50$\pm$ 0.50 & 0.294E+08$\pm$0.255E+07 &  0.82$\pm$ 0.06 \\
170009 & 10500$\pm$ 600 & 3.37$\pm$0.17 &  2.68$\pm$ 0.16 &  3.80$\pm$ 0.30 & 0.228E+09$\pm$0.208E+08 &  1.01$\pm$ 0.03 \\
170714 & 23500$\pm$1300 & 4.03$\pm$0.16 &  3.81$\pm$ 0.12 &  9.80$\pm$ 0.70 & 0.117E+08$\pm$0.350E+07 &  0.42$\pm$ 0.12 \\
171219 & 15500$\pm$ 800 & 3.62$\pm$0.14 &  3.30$\pm$ 0.12 &  6.00$\pm$ 0.30 & 0.644E+08$\pm$0.500E+07 &  0.88$\pm$ 0.05 \\
173219 & 28000$\pm$2200 & 3.88$\pm$0.21 &  4.45$\pm$ 0.16 & 14.80$\pm$ 1.60 & 0.892E+07$\pm$0.211E+07 &  0.60$\pm$ 0.13 \\
173371 & 12500$\pm$ 600 & 3.72$\pm$0.10 &  2.69$\pm$ 0.08 &  4.30$\pm$ 0.20 & 0.138E+09$\pm$0.943E+07 &  0.81$\pm$ 0.04 \\
173530 & 12500$\pm$1100 & 3.25$\pm$0.18 &  3.23$\pm$ 0.10 &  5.20$\pm$ 0.30 & 0.108E+09$\pm$0.808E+07 &  1.02$\pm$ 0.02 \\
173637 & 28000$\pm$1700 & 3.85$\pm$0.15 &  4.45$\pm$ 0.09 & 14.60$\pm$ 1.10 & 0.958E+07$\pm$0.154E+07 &  0.64$\pm$ 0.09 \\
173817 & 13500$\pm$1100 & 3.78$\pm$0.19 &  2.74$\pm$ 0.12 &  4.50$\pm$ 0.30 & 0.116E+09$\pm$0.177E+08 &  0.76$\pm$ 0.09 \\
174513 & 23000$\pm$1700 & 3.95$\pm$0.17 &  3.86$\pm$ 0.10 &  9.80$\pm$ 0.80 & 0.151E+08$\pm$0.401E+07 &  0.54$\pm$ 0.13 \\
174571 & 21500$\pm$1700 & 4.25$\pm$0.24 &  3.36$\pm$ 0.21 &  8.00$\pm$ 0.90 & 0.249E+07$\pm$0.456E+07 &  0.06$\pm$ 0.14 \\
174705 & 22500$\pm$1700 & 4.08$\pm$0.17 &  3.67$\pm$ 0.09 &  9.10$\pm$ 0.70 & 0.108E+08$\pm$0.459E+07 &  0.34$\pm$ 0.13 \\
174886 & 14500$\pm$ 600 & 3.40$\pm$0.10 &  3.34$\pm$ 0.08 &  5.70$\pm$ 0.30 & 0.829E+08$\pm$0.517E+07 &  1.00$\pm$ 0.02 \\
175869 & 11500$\pm$ 600 & 3.31$\pm$0.11 &  2.91$\pm$ 0.09 &  4.30$\pm$ 0.20 & 0.170E+09$\pm$0.102E+08 &  1.02$\pm$ 0.01 \\
176159 & 13500$\pm$1100 & 3.73$\pm$0.18 &  2.84$\pm$ 0.10 &  4.70$\pm$ 0.30 & 0.109E+09$\pm$0.151E+08 &  0.80$\pm$ 0.08 \\
176630 & 16000$\pm$1100 & 3.29$\pm$0.16 &  3.76$\pm$ 0.13 &  7.30$\pm$ 0.50 & 0.492E+08$\pm$0.368E+07 &  1.02$\pm$ 0.02 \\
178479 & 16500$\pm$1100 & 4.00$\pm$0.16 &  2.94$\pm$ 0.09 &  5.50$\pm$ 0.30 & 0.466E+08$\pm$0.131E+08 &  0.50$\pm$ 0.12 \\
179343 & 11000$\pm$1100 & 2.90$\pm$0.22 &  3.34$\pm$ 0.18 &  5.50$\pm$ 0.60 & 0.925E+08$\pm$0.125E+08 &  1.02$\pm$ 0.00 \\
  179405 & 20000$\pm$2200 & 4.25$\pm$0.27 & 3.15$\pm$0.16 & 6.99$\pm$0.76 &  0.134E+08$\pm$0.588E+07 & 0.21$\pm$0.09\\
180126 & 20500$\pm$1700 & 4.01$\pm$0.19 &  3.48$\pm$ 0.10 &  7.80$\pm$ 0.60 & 0.197E+08$\pm$0.670E+07 &  0.45$\pm$ 0.13 \\
181231 & 13500$\pm$1100 & 3.76$\pm$0.18 &  2.79$\pm$ 0.10 &  4.60$\pm$ 0.30 & 0.110E+09$\pm$0.162E+08 &  0.77$\pm$ 0.08 \\
181308 & 14000$\pm$1100 & 3.84$\pm$0.17 &  2.81$\pm$ 0.10 &  4.70$\pm$ 0.30 & 0.914E+08$\pm$0.160E+08 &  0.70$\pm$ 0.10 \\
  181367 & 15000$\pm$1100 & 3.76$\pm$0.17 & 2.99$\pm$0.12 & 5.21$\pm$0.32 &  0.795E+08$\pm$0.808E+07 & 0.75$\pm$0.06\\
181709 & 13500$\pm$1100 & 3.32$\pm$0.20 &  3.31$\pm$ 0.17 &  5.50$\pm$ 0.50 & 0.942E+08$\pm$0.927E+07 &  1.02$\pm$ 0.03 \\
181803 & 12000$\pm$ 600 & 3.03$\pm$0.18 &  3.38$\pm$ 0.22 &  5.60$\pm$ 0.80 & 0.875E+08$\pm$0.143E+08 &  1.02$\pm$ 0.00 \\
184279 & 29500$\pm$2200 & 4.00$\pm$0.22 &  4.43$\pm$ 0.18 & 15.50$\pm$ 1.80 & 0.610E+07$\pm$0.209E+07 &  0.43$\pm$ 0.14 \\
184767 &  9500$\pm$ 600 & 3.46$\pm$0.13 &  2.37$\pm$ 0.08 &  3.30$\pm$ 0.20 & 0.323E+09$\pm$0.266E+08 &  0.98$\pm$ 0.03 \\
194244 & 10000$\pm$ 600 & 3.35$\pm$0.12 &  2.62$\pm$ 0.07 &  3.70$\pm$ 0.10 & 0.259E+09$\pm$0.150E+08 &  1.02$\pm$ 0.02 \\
  230579 & 30000$\pm$1700 & 4.00$\pm$0.16 & 4.49$\pm$0.13 &16.40$\pm$1.33 &  0.535E+07$\pm$0.114E+07 & 0.43$\pm$0.09\\
BD-09\d4858 & 21000$\pm$1700 & 4.25$\pm$0.19 &  3.29$\pm$ 0.12 &  7.60$\pm$ 0.70 & 0.258E+07$\pm$0.480E+07 &  0.06$\pm$ 0.11 \\
\hline
\end{longtable}

\end{document}